\documentclass[fleqn,usenatbib]{mnras}

\usepackage{newtxtext,newtxmath}

\usepackage[T1]{fontenc}
\usepackage{ae,aecompl}

\usepackage{pdflscape}
\usepackage{pdfpages}
\usepackage{adjustbox}
\usepackage{multirow}
\usepackage[usestackEOL]{stackengine}

\usepackage{graphicx}	
\usepackage{amsmath}	
\usepackage{ulem}

\newcommand{\kms}{\,km\,s$^{-1}$}
\newcommand{\pab}{Pa $\rm \beta1.28$}
\newcommand{\afe}{$[\alpha/{\rm Fe}]$}
\newcommand{\ms}{M$_{\sun}$}

\title[NIR spectral Indices]{ Fingerprints of Stellar Populations in the Near-Infrared: An Optimised Set of Spectral Indices in the \textit{JHK}-Bands}

\author[E. Eftekhari et al.]{
Elham Eftekhari,$^{1,2}$\thanks{E-mail: elhamea@iac.es }
Alexandre Vazdekis,$^{1,2}$
and Francesco La Barbera$^{3}$
\\
$^{1}$Instituto de Astrof\'isica de Canarias, E-38200 La Laguna, Tenerife, Spain\\
$^{2}$Departamento de Astrof\'isica, Universidad de La Laguna, E-38205 La Laguna, Tenerife, Spain\\
$^{3}$INAF-Osservatorio Astronomico di Capodimonte, sal. Moiariello 16, Napoli I-80131, Italy\\
}

\date{Accepted 2021 April 1. Received 2021 February 28; in original form 2020 August 22}

\pubyear{2020}

\begin{document}
\label{firstpage}
\pagerange{\pageref{firstpage}--\pageref{lastpage}}
\maketitle

\begin{abstract}

Stellar population studies provide unique clues to constrain galaxy formation models. So far, detailed studies based on absorption line strengths have mainly focused on the optical spectral range although many diagnostic features are present in other spectral windows. In particular, the near-infrared (NIR) can provide a wealth of information about stars, such as evolved giants, that have less evident optical signatures. Due to significant advances in NIR instrumentation and extension of spectral libraries and stellar population synthesis (SPS) models to this domain, it is now possible to perform in-depth studies of spectral features in the NIR to a high level of precision. In the present work, taking advantage of state-of-the-art SPS models covering the NIR spectral range, we introduce a new set of NIR indices constructed to be maximally sensitive to the main stellar population parameters, namely age, metallicity and initial mass function (IMF). We fully characterize the new indices against these parameters as well as their sensitivity to individual elemental abundance variations, velocity dispersion broadening, wavelength shifts, signal-to-noise ratio and flux calibration. We also present, for the first time, a method to ensure that the analysis of spectral indices is not affected by sky contamination, which is a major challenge when dealing with NIR spectroscopy. Moreover, we discuss two main applications: (i) the ability of some NIR spectral indices to constrain the shape of the low-mass IMF and (ii) current issues in the analysis of NIR spectral indices for future developments of SPS modelling.

\end{abstract}




\section{Introduction}

Understanding the evolutionary history of galaxies has shed light on many physical processes such as supernova explosions and gamma-ray burst events, evolution and merger of supermassive black holes, dust formation and destruction, stellar dynamics, interstellar medium enrichment and even histories of planet formation. Hence, a better picture of galaxy formation and evolution is beneficial to many other areas of astrophysics. One way to achieve this goal is by studying the stellar content of galaxies since it provides valuable information on different star formation episodes in a galaxy and its chemical evolution. 

Spectroscopy-based studies of the stellar population content of galaxies are based on the comparison between evolutionary stellar population synthesis (SPS) models and data. This comparison can be made by using either line-strength fitting \citep[e.g.][]{worthey1994, vazdekis1997, trager2000, thomas2005, gallazzi2008, riffel2008, graves2009, labarbera2013, spiniello2014, martin2015b, martin2015d, martin2015c, labarbera2015, labarbera2016, labarbera2017, rosani2018, labarbera2019, ferreras2019, salvador2019, salvador2021}, full spectral fitting \citep[e.g.][]{vazdekis1999a, vazdekis1999b, riffel2009, riffel2010, conroy2012b, ferre2013, labarbera2014, choi2014, vandokkum2017}, or full index fitting (FIF) \citet{martin2019} methods. In the line-strength approach, the stellar population parameters of galaxies are determined by fitting the pseudo-equivalent width of spectral indices, while in full spectral fitting, fluxes at different wavelengths are fitted simultaneously. The FIF is a new approach, and it uses index definitions to normalize the continuum and then fits pixels within the feature. Each method has its own advantages and drawbacks. For example, the advantage of using line-strength indices is that the focus is given only to relatively narrow and well characterized spectral regions particularly sensitive to relevant stellar population parameters. However, this advantage is achieved with a lots of information from all pixels in the spectrum and also in the spectral shape of the feature. Moreover, in line-strength fitting, the absorption line-strengths are measured with respect to pseudo-continua, which in reality are blends of several element species. Therefore, variation of an index may be due to a combination of variations in the feature itself and to variations in the pseudo-continua, which are due to different stellar population parameters. On the other hand, spectral fitting can use the entire measured signal rather than partial information encoded in the spectral indices. This approach involves many free parameters, not necessarily well constrained, and demands much higher model accuracy. It assumes, ideally, that model predictions are 'perfect', and all systematics from data reduction are fully under control. The advantage of the hybrid approach (FIF) is that it focuses on narrow spectral regions, as in the line-strength method, trying to maximize the information we can extract from the spectrum in those regions.

Although the information about the properties of a stellar population is distributed over the entire spectrum, this information is highly redundant. Hence, regardless of the technique of fitting models to data, it is important to have a system of spectral indices, in order to extract relevant information required for the stellar population analysis. So far, detailed analysis of the spectral absorption features in the spectrum of galaxies has been concentrated mainly on the optical wavelengths. Diagnostics of the stellar content of galaxies have been identified and well-characterized thanks to high-quality empirical spectral libraries in the optical SPS models \citep[e.g.][]{worthey1994, vazdekis1996, vazdekis1999a, bruzual2003, thomas2003, leborgne2004, maraston2005, vazdekis2010, maraston2011, conroy2012a, vazdekis2012, vazdekis2016, conroy2018, maraston2020}.  The Lick IDS instrument dependent system of indices  \citep{worthey1994} and its flux-scaled LIS version \citep{vazdekis2010} is the most widely used set of indices in the optical that contains the definition (wavelength limits) of these diagnostic features. 

Due to the challenges of spectroscopy with ground-based telescopes in wavelengths beyond 9000 \AA, the near-infrared (NIR) range has been relatively unexplored. The background radiation is significant, and absorption from the Earth's atmosphere is strong in the NIR. These effects make observational strategies and data reduction in the NIR domain complicated. However, spectroscopy at NIR wavelengths has the advantage of probing optically-obscured regions of galaxies. Moreover, cool evolved stars such as short-lived red supergiants (in very young populations $\sim 10$ Myr) and thermally-pulsating asymptotic giant branch (TP-AGB) stars (in intermediate-age populations 0.1-2 Gyr) or red giant branch stars (in old stellar populations) make a strong contribution to the total flux of galaxies in the NIR. However, the NIR domain has lower sensitivity to the young stellar component of galaxies that dominate the optical part of the spectrum. Hence, NIR spectroscopy allows the degeneracy between multiple stellar populations to be broken, providing information complementary to the optical range.  

Owing to the advent of NIR instrumentation, it has recently become possible to measure absorption features in the NIR with the accuracy required for stellar population studies. In parallel to the development of NIR instruments, theoretical and empirical spectral libraries have been extended to this window (IRTF \citet{rayner2009}, XSL \citet{gonneau2020}) and used as a basis for the population synthesis models in the NIR (\citet{maraston2005}, \citet{conroy2012a}, \citet{meneses2015}, E-MILES \citet{rock2015, vazdekis2016, rock2016}, A-LIST \citet{ashok2021}). Unlike in the optical range, there is no homogeneous, optimized and fully characterized set of indices, covering the \textit{J}-, \textit{H}- and \textit{K}-bands, to compare the predictions of spectral synthesis models to observations. In the last two decades, several studies have attempted to utilise NIR absorption features in the study of stellar populations. These studies have either identified new spectral indices or modified previous definitions, based on different criteria. In many cases only a narrow spectral window was covered. For example, \citet{frogel2001} modified some of the wavelength intervals of only three spectral indices at the end of the \textit{K}-band (\ion{Na}{i}, \ion{Ca}{i}, and CO). \citet{ivanov2004} defined new indices for atomic lines and also adopted previous definitions for some indices in the \textit{H}- and \textit{K}-bands. \citet{silva2008} considered velocity broadening of galaxy spectra and optimised the definitions of \ion{Fe}{i} and \ion{Mg}{i} absorptions at the red end of the \textit{K}-band. \citet{marmol2008} optimised the CO index at 2.3 $\mu$m. CvD12 supplemented the blue Lick indices with new indices in the NIR to constrain the number of low mass stars in the spectrum of galaxies. \citet{rock2015phd} proposed a homogeneous set of indices in the NIR. The author re-defined a number of  indices in the \textit{J}-, \textit{H}- and \textit{K}-bands based on E-MILES SPS models and characterised them as a function of age, metallicity and IMF slope. \citet{riffel2019} presented new definitions for the NIR absorption features, taking into account the location of the most common emission lines detected in active galaxies. \citet{cesetti2013} and \citet{morelli2020} presented an extensive study of absorption features in the \textit{I}, \textit{Y}, \textit{J}, \textit{H}, \textit{K} and \textit{L} windows. They optimised the definition of indices to be sensitive to physical stellar parameters, i.e. $\rm T_{eff}$, [Fe/H], and log(g) and they characterized the dependence of their indices on velocity dispersion broadening. However, these studies lacked an optimization, relevant stellar population parameters wise, and full characterization of each index (see below). Moreover, the existence of various definitions for the same absorption in the literature makes the comparison of results from different studies far from trivial. For these reasons, in the present work, we provide a new set of optimized NIR indices, fully characterized for stellar population analysis. The indices are characterized not only with respect to the main physical properties of galaxies, namely age, metallicity and stellar initial mass function (IMF) but also with respect to the effect of elemental abundance ratios, velocity dispersion, wavelength shifts, signal-to-noise ratio (SNR), flux calibration and the contamination from sky emission/absorption lines. For the first time, we also describe a procedure to select indices that can be robustly used for stellar population analysis in the NIR.

The layout of the paper is the following. In Sect.~\ref{sec:absorption_identification}, we present the method to identify the most sensitive absorption features to a particular stellar population parameter (age, metallicity and IMF). The procedure to define spectral indices is described in Sect.~\ref{sec:optimised_index_definition}, while a thorough characterization of the indices is provided in Sect.~\ref{sec:characterization}. In Sect.~\ref{sec:example_applications}, we describe practical applications of our newly defined indices. The paper closes with a summary in Sect.~\ref{sec:summary}.

\section{Absorption Identification} \label{sec:absorption_identification}

We aim to determine which absorptions in the NIR spectra of unresolved populations could be of interest for studying their stellar content. Simple stellar population (SSP) spectra provide us with a flexible tool for such kind of analysis. In this work, we used E-MILES SSPs \citep{vazdekis2016} which adopt the IRTF stellar library \citep{cushing2005, rayner2009} in the NIR. The E-MILES SSPs cover a wide range of ages, from 1 to 14 Gyr (1 to 17.78 Gyr), and metallicity between -0.35 to +0.26 dex (-0.4 to +0.22 dex), for models based on BaSTI isochrones \citep{pietrinferni2004} (Padova00 isochrones \citep{girardi2000}). They are computed for a variety of IMFs, including both single power-law (unimodal) and low-mass tapered (bimodal) IMF, whose slope $\Gamma$ and  $\Gamma_{b}$, respectively, increases as the IMF becomes more enriched in dwarfs relative to giant stars  (i.e. more bottom-heavy compared to a Kroupa-like distribution, see \citet{vazdekis1996, vazdekis1997}, for details). To identify the most sensitive spectral features to a given stellar population parameter, we computed "response functions" of SSP models to age, metallicity and IMF slope, respectively. Each response is obtained by dividing a given SSP spectrum to a reference model. For the response to age, we divided two SSPs of solar metallicity and Kroupa-like IMF with different ages, i.e. 12 and 2 Gyr, respectively. To obtain the metallicity response function, we divided SSP spectra with the same IMF (Kroupa-like) and the same age (12 Gyr) but with [M/H] = +0.15 and -0.25 dex, respectively. Finally, for the IMF sensitivity, we divided an SSP with bottom-heavy bimodal IMF (slope $\Gamma_{b}=3$; which is typical for the central regions of massive early-type galaxies (ETGs), see \citet{labarbera2013}) by an SSP with Kroupa-like IMF. Both models have solar metallicity and 12 Gyr age. All SSPs used to compute response functions are base models (see, e.g., \citet{vazdekis2015}), computed with BaSTI isochrones and broadened to $\sigma = 360$\kms \footnote{This is a conservative choice as it is the typical velocity dispersion of highest mass galaxies, corresponding to the "maximum" smoothing one might have when studying unresolved stellar populations.}.

\begin{figure*}
	\includegraphics[width=0.89\linewidth]{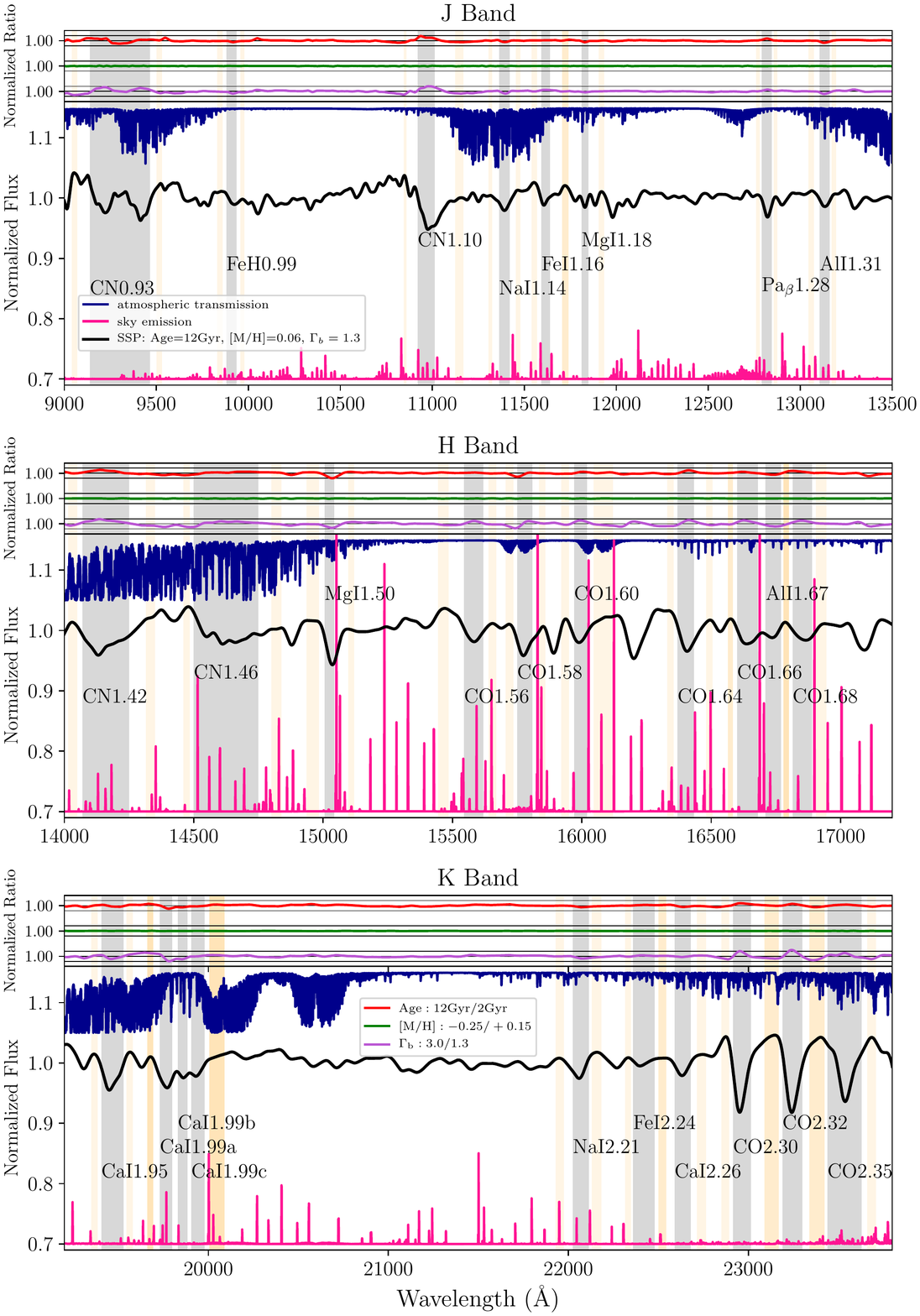}
    \caption{The plots show  \textit{J}- (top), \textit{H}- (middle), and \textit{K}- (bottom) band spectral regions, respectively. The top panel of each spectral region shows the flux ratios obtained by dividing two models that differ only in the adopted parameter (age, metallicity and IMF slope; plotted with red, green, and purple colours (see the inset in the bottom plot)). Grey and orange bands represent absorption and continuum bandpasses, for each index definition, overplotted on a spectrum of E-MILES SSP model (black). A scaled telluric absorption spectrum (dark blue) and sky emission spectrum (pink) is shown to better highlight clear sky regions (see the inset in the upper plot). }
    \label{fig:fig1}
\end{figure*}

To create response functions to a given parameter, we removed the effect of different continuum shapes between different SSPs, by fitting second-degree splines to the SSPs ratios. The sensitivities obtained in this way, provide us with useful visual guidelines for identifying potential proxies. In the top panel of each plot in Fig.~\ref{fig:fig1}, we show the SSPs response functions. The red, green and purple lines show age, metallicity and IMF sensitivities, respectively. The upper and lower black lines, around each response function, show $\pm 1\%$ sensitivities. 

As our reference spectrum in defining NIR indices, we used an SSP of solar metallicity, Kroupa-like IMF and 12 Gyr age. This spectrum is shown with a thick black line in the bottom panels of each plot in Fig.~\ref{fig:fig1}. In the same panels, we also show (arbitrarely rescaled) sky emission lines in pink and telluric absorption lines in dark blue (both obtained by the ESO Skycalc tool \citep{noll2012, jones2013}).  

By looking at the SSP spectral ratios, we find regions that are more sensitive to a given stellar population parameter. In order to identify which atomic/molecular species are responsible for the corresponding absorption features in the reference spectrum, we proceed as follows. Since the IRTF library is used to build up the E-MILES models in the NIR, we used the prominent Arcturus atomic lines identified in this library (table 7 of \citet{rayner2009}). Moreover, we used molecular band identifications in the IRTF library from table 10 of \citet{rayner2009}. We also made use of M-dwarf absorption features which are identified as IMF-sensitive features in \cite{lagattuta2017}, as well as line lists from \citet{kleinmann1986}. Appendix~\ref{sec:appendixA} shows a zoom in of each index and indicates identified element species.

Figure~\ref{fig:fig1} shows that the bluest part of the \textit{J}-Band contains CN and FeH absorptions at about $9000-10000$ \AA. The sensitivities around $11000-12000$ \AA \space show rises and dips with variable strength, corresponding to CN, \ion{Na}{i}, \ion{Fe}{i} and \ion{Mg}{i} features, respectively. Pa$\rm \beta$ and \ion{Al}{i} are also indicated with a peak and a dip, respectively, in both age and IMF sensitivities. CO molecular absorptions at about $15000-17000$ \AA \space are the most prominent features in \textit{H}-band. The CN molecules and metallic species such as \ion{Mg}{i} and \ion{Al}{i} are also present in the \textit{H}-band. In the \textit{K}-band, the sensitivity of IMF shows high peaks beyond $23000$ \AA \space corresponding to CO band heads. \ion{Ca}{i} lines correspond to the dips in sensitivities in the bluest part of the \textit{K}-band. Sensitivity to metallicity is almost flat in the entire NIR spectral range.

\section{Optimised Index Definition} \label{sec:optimised_index_definition}

To define a key set of spectral indices in the NIR, we use a Lick-style index definition, where each index consists of a central bandpass enclosing the feature, as well as a blue and red pseudo-continuum bandpasses at either side of the feature. The SSP response functions provide us with a useful tool to define optimised central wavelengths and pseudo-continuum bands with respect to stellar population parameters. Indeed, we included the peak of sensitivities for selecting central bandpass of indices if associated with an absorption. For defining spectral indices, we also take into account several criteria, as described in the following.

Atmospheric absorption is a severe challenge in the NIR, as several indices are affected by telluric lines. The strongest atmospheric absorptions in the NIR are water vapour features, while atmospheric transmission is also affected by $\rm O_{2}$, $\rm CO_{2}$ and $\rm CH_{4}$. Sky emission lines, especially in the \textit{H}-band, are another source of concern when studying NIR spectral features. We tried not to include strong telluric absorption and sky emission lines in the wavelength range of the index definition; however, it was almost unavoidable. Note that we are aided in this task by the E-MILES models, which are fed with empirical stellar spectra.  Hence, our effort to define indices with minimum contamination from sky lines helps to obtain clean SPS model predictions (as the models are based on the spectra of nearby stars), as well as measure clean indices for galaxies at redshift z$\sim 0$. Since the effect of sky residuals might be confused with that of stellar population parameters, in Sect.~\ref{sec:reliability} we also introduce a more general methodology to recognize indices that cannot be used when  analysing a given set of galaxy spectra in the NIR.

In order to have indices that are robust against systematic effects arising in the flux calibration, the wavelength range involved should be quite narrow. Hence, we tried not to put pseudo-continuum bands very far apart from the absorption feature. Indeed, $65\%$ of our indices are less than 300 \AA \space wide. This means that most of our indices are robust against flux calibration with uncertainties affecting wavelength scales of (typically) $\sim$300 \AA.

Absorption features broaden due to galaxy velocity dispersion or instrumental resolution. Line-strength might be affected if the pseudo-continua overlaps with the feature bandpass \citep{vazdekis1999b, vazdekis2001}. To ensure that pseudo-continua are not going to be contaminated by the smeared absorption feature, we have defined sideband limits to be at least 10 \AA \space far from the central band.

The widths of bandpasses were chosen to minimise the effect of typical galaxy velocity dispersions on the line-strengths. As the bandpasses become narrower, the sensitivity of the index to velocity dispersion becomes generally higher, but the bandpass width should not be too broad; since it can be affected by metallic lines. In order to be conservative, we put a minimum width of 10 \AA \space for bandpasses, since narrower definitions might be sensitive to the Poissonian noise at scales shorter than velocity dispersion \citep{cardiel1998}. 

We changed the position of central and pseudo-continua bandpasses or modified their widths on trial index definitions until optimizing the definition of indices to the above requirements. We defined spectral indices as presented in Table~\ref{tab:tab1}. At the effective resolutions imposed by galaxy dynamics, none of these atomic and molecular lines is contributed by a single element. Column 8 reports the elements whose absorption were found to contribute to the given feature, with the feature name (Col. 1) coming from what we identified as the possible main contributor to the absorption\footnote{Traditionally, the nomenclature of spectral indices originates from the main element contributor to the absorption feature. Since in  \citet{rayner2009}, the depth of strong Arcturus metal lines  (detectable in the IRTF stellar library) has not been provided, we identified the main contributor to the absorption feature by considering the number of absorption lines arising from one element, the response of the line-strength index to the variation of elemental abundances, and identification of the main contributor to the feature from the literature (see Appendix~\ref{sec:appendixA}, when available).}. Columns 2 to 7 show the limiting wavelengths of bandpasses. All wavelengths in this paper are quoted in the air system. Some of these indices were already defined and studied in other works (e.g. \ion{Na}{i}1.14, \ion{Mg}{i}1.50, \ion{Na}{i}2.21, CO2.30, etc.). In this paper, we re-defined and optimised them according to our criteria. Appendix~\ref{sec:appendixA} compares all the definitions of the same feature that we found in the literature.

Figure~\ref{fig:fig1} shows our newly defined indices. The grey and orange regions mark the bandpasses of the features and their adjacent pseudo-continua, respectively.

\begin{table*}
	\centering
	\caption{Definition of spectral indices. Column 1 gives the name of indices. Definition of blue band edges is provided in Cols. 2 and 3. The wavelength limits of the central bandpass are given in Cols. 4 and 5. Columns 6 and 7 provide the definition of red band edges. Column 8 lists atomic and molecular species that contribute to the index. All wavelengths are quoted in the air system.}
	\begin{adjustbox}{width=\textwidth}
	\label{tab:tab1}
	\begin{tabular}{c c c c c c c c c} 
		\hline\hline
		Index	& $\rm \lambda_{blue1}$	& $\rm \lambda_{blue2}$	& $\rm \lambda_{centre1}$ &	$\rm \lambda_{centre2}$	& $\rm \lambda_{red1}$	& $\rm \lambda_{red2}$	& Absorber(s)	\\
		 & (\AA)	& (\AA)	& (\AA) &	(\AA)	& (\AA)	& (\AA)	& 	\\
		(1)	& (2)	& (3)	& (4) &	(5)	& (6)	& (7)	& (8)	\\
		\hline
CN0.93 & 9040.0 & 9070.0 & 9138.0 & 9465.0 & 9500.0 & 9530.0 & CN/ZrO/TiO/\ion{Ti}{i}/\ion{Si}{i}/\ion{Fe}{i}/\ion{Mg}{i}/\ion{Cr}{i} \\
FeH0.99 & 9830.0 & 9860.0 & 9880.0 & 9935.0 & 9955.0 & 9980.0 & FeH/\ion{Fe}{i}/\ion{Ti}{i}/TiO \\
CN1.10 & 10845.0 & 10860.0 & 10920.0 & 11012.0 & 11124.0 & 11170.0 & CN/\ion{Mg}{i}/\ion{Si}{i}/\ion{Fe}{i}/\ion{Ni}{i} \\
\ion{Na}{i}1.14 & 11305.0 & 11325.0 & 11362.0 & 11420.0 & 11453.0 & 11480.0 & \ion{Na}{i}/\ion{Cr}{i}/\ion{Fe}{i} \\
\ion{Fe}{i}1.16 & 11540.0 & 11570.0 & 11593.0 & 11640.0 & 11705.0 & 11740.0 & \ion{Fe}{i}/\ion{Si}{i}/\ion{Cr}{i} \\
\ion{Mg}{i}1.18 & 11705.0 & 11740.0 & 11810.0 & 11850.0 & 11905.0 & 11935.0 & \ion{Mg}{i}/\ion{Ca}{ii} \\
\pab & 12760.0 & 12780.0 & 12790.0 & 12845.0 & 12855.0 & 12875.0 & Pa $\rm \beta$/\ion{Fe}{i}/\ion{Ti}{i}/\ion{Ca}{i} \\
\ion{Al}{i}1.31 & 13045.0 & 13075.0 & 13105.0 & 13160.0 & 13170.0 & 13195.0 & \ion{Al}{i}/\ion{Ca}{i}/\ion{Fe}{i} \\
CN1.42 & 14015.0 & 14050.0 & 14070.0 & 14250.0 & 14315.0 & 14350.0 & CN \\
CN1.46 & 14460.0 & 14485.0 & 14500.0 & 14750.0 & 14800.0 & 14840.0 & CN/VO \\
\ion{Mg}{i}1.50 & 14935.0 & 14985.0 & 15005.0 & 15044.0 & 15097.0 & 15120.0 & \ion{Mg}{i}/\ion{Fe}{i} \\
CO1.56 & 15445.0 & 15485.0 & 15545.0 & 15620.0 & 15640.0 & 15670.0 & $^{12}$CO/\ion{Fe}{i}/\ion{Si}{i}/\ion{Ti}{i}/\ion{Ni}{i} \\
CO1.58 & 15705.0 & 15735.0 & 15750.0 & 15810.0 & 15835.0 & 15860.0 & $^{12}$CO/\ion{Fe}{i}/\ion{Mg}{i} \\
CO1.60 & 15920.0 & 15950.0 & 15970.0 & 16020.0 & 16045.0 & 16120.0 & $^{12}$CO/\ion{Fe}{i} \\
CO1.64 & 16330.0 & 16355.0 & 16370.0 & 16434.0 & 16480.0 & 16505.0 & $^{12}$CO/\ion{Fe}{i}/\ion{Si}{i} \\
CO1.66 & 16565.0 & 16585.0 & 16600.0 & 16680.0 & 16780.0 & 16800.0 & $^{12}$CO/\ion{Fe}{i}/\ion{Si}{i} \\
\ion{Al}{i}1.67 & 16565.0 & 16585.0 & 16710.0 & 16770.0 & 16780.0 & 16800.0 & \ion{Al}{i}/\ion{Fe}{i} \\
CO1.68 & 16780.0 & 16800.0 & 16815.0 & 16890.0 & 16905.0 & 16945.0 & $^{12}$CO/\ion{Si}{i}/\ion{Fe}{i}/\ion{Ni}{i} \\
\ion{Ca}{i}1.95 & 19350.0 & 19385.0 & 19405.0 & 19530.0 & 19545.0 & 19580.0 & \ion{Ca}{i}/\ion{Mg}{i}/\ion{Si}{i}/\ion{Fe}{i} \\
\ion{Ca}{i}1.99a & 19660.0 & 19695.0 & 19730.0 & 19800.0 & 20005.0 & 20090.0 & \ion{Ca}{i}/\ion{Mg}{i}/\ion{Fe}{i} \\
\ion{Ca}{i}1.99b & 19660.0 & 19695.0 & 19830.0 & 19885.0 & 20005.0 & 20090.0 & \ion{Ca}{i}/\ion{Fe}{i} \\
\ion{Ca}{i}1.99c & 19660.0 & 19695.0 & 19905.0 & 19980.0 & 20005.0 & 20090.0 & \ion{Ca}{i}/\ion{Fe}{i}/\ion{Si}{i} \\
\ion{Na}{i}2.21 & 21930.0 & 21975.0 & 22025.0 & 22114.0 & 22130.0 & 22185.0 & \ion{Na}{i}/\ion{Si}{i}/ScI \\
\ion{Fe}{i}2.24 & 22315.0 & 22350.0 & 22360.0 & 22480.0 & 22500.0 & 22545.0 & \ion{Fe}{i}/\ion{Ti}{i} \\
\ion{Ca}{i}2.26 & 22500.0 & 22545.0 & 22590.0 & 22680.0 & 22715.0 & 22765.0 & \ion{Ca}{i} \\
CO2.30 & 22850.0 & 22895.0 & 22915.0 & 23015.0 & 23090.0 & 23170.0 & $^{12}$CO \\
CO2.32 & 23090.0 & 23170.0 & 23190.0 & 23300.0 & 23340.0 & 23425.0 & $^{12}$CO \\
CO2.35 & 23340.0 & 23425.0 & 23440.0 & 23630.0 & 23660.0 & 23710.0 & $^{12}$CO \\
		\hline
		
	\end{tabular}
	\end{adjustbox}
\end{table*}

\section{Characterization of Indices} \label{sec:characterization}

Spectral population synthesis allows us to measure absorption features of SSP models with given metallicity, age, stellar IMF, and elemental abundance. In this section, we utilize base E-MILES models, as well as \citet{conroy2012a} (hereafter CvD12) models to study the behaviour of our new set of spectral indices extensively. Since E-MILES models do not include the effect of individual abundance ratios, we use CvD12 models to evaluate the sensitivity of indices to individual abundance variations, while relying on E-MILES SSPs to characterize the dependence of the indices on the main stellar populations parameters (age, metallicity, and IMF slope). The use of SSP models does not only allow us to characterize spectral indices as a function of population parameters but also characterize their sensitivity to the Doppler broadening, SNR requirements, as well as wavelength shifts due to, e.g., kinematic rotation, recessional velocity and uncertainties due to wavelength calibration, as detailed in the following.

\subsection{Index behaviour as a function of age, metallicity and IMF}\label{sec:age_metallicity_imf}

The characterization of the line-strength indices defined in this work is shown in Figs.~\ref{fig:fig2} to \ref{fig:fig6}. The panels in Col. 'a' of all figures show the sensitivity of indices to age. The indices are measured on the four different sets of base E-MILES models: Two sets of models with bimodal IMF of slope 1.3, which is representative of the Milky Way IMF, with solar and super-solar metallicities (green and black solid lines) and two other sets of models with bimodal IMF of slope 3.0, representative of bottom-heavy IMF in massive galaxies, with solar and super-solar metallicities (green and black dashed lines). Each set of models span a range in age from 1 to 14 Gyr. These panels show that indices redward 15000 \AA \space stay almost flat for ages greater than 2 Gyr. Exceptions are calcium indices and the sodium index in \textit{K}-band. Comparing green and black colours shows that the dependence of NIR indices on age does not change with metallicity. Only for bottom-heavy models and old ages, the CN0.93, CN1.42 and CN1.46 indices show some differences in time evolution for different metallicities. It is also noteworthy to mention that the line-strength of Pa $\rm\beta$ index at 1.28 \micron \space (Fig.~\ref{fig:fig3}) drops sharply with age for populations younger than 2 Gyr. The comparison of solid lines with dashed ones reveals that except \ion{Al}{i}1.67 in \textit{H}-band, indices in the NIR have a mild to significant dependence on IMF. For instance, the larger line-strengths of CO and CN indices in SSPs with IMF slope of 1.3 (solid lines) with respect to  SSPs with IMF slope of 3.0 (dashed lines) are noticeable.

The AGB evolutionary phase has essential contributions to the NIR light of stellar populations with intermediate age. According to \citet{maraston2005}, this phase is dominated by the TP-AGB stars. They show that inclusion of the TP-AGB stars in the modelling of 0.3 $\lesssim$ t $\lesssim$ 2 Gyr stellar populations has a substantial impact not only on the absolute flux but also on the absorption features such as CN, $\rm C_{2}$, $\rm H_{2}O$ and $\rm CO_{2}$. Therefore, the peak at around 2 Gyr in the CN0.93, CN1.10, CN1.42 and CN1.46 indices in Col. 'a' of Figs.~\ref{fig:fig2} and \ref{fig:fig3} is an indicator of intermediate-age stellar populations in the integrated light of galaxies. The first detection of the CN molecular band at $\sim$11000 \AA \space was reported by \citet{riffel2007} in a sample of Seyfert galaxies, implying the presence of recent star formation episodes in these systems. 

The panels in Col. 'b' of Figs.~\ref{fig:fig2} to \ref{fig:fig6} show the line-strengths of NIR indices as a function of metallicity, for E-MILES SSPs with a fixed Kroupa-like IMF, and young (2 Gyr; blue lines) and old (12 Gyr; red lines) populations. The total metallicity in each set of spectra varies from -0.35 to +0.26 dex. According to these panels, CN indices in \textit{J}- and \textit{H}-bands are the most sensitive ones to the metallicity. FeH0.99, \ion{Al}{i}1.31, CO1.60, CO1.64 and \ion{Ca}{i}1.95 show a very mild trend with metallicity.  The strength of CN0.93 decreases rapidly from solar to super-solar metallicity. CN indices, \ion{Na}{i}1.14, \pab \space and CO indices in \textit{K}-band have different sensitivity to metallicity when comparing  young to  old populations (i.e. comparing blue to red lines), while other indices do not change significantly. Note that the line-strength of most of the indices varies in a non-linear way with metallicity for [M/H] $\geq +0.06$ dex, especially in young populations (solid blue lines). For instance, in Fig.~\ref{fig:fig3}, for young populations (blue line), \pab \space decreases as a function of total metallicity in the range $\rm -0.25 \leq [M/H] \leq 0.06$, but then it increases and goes down again at the highest [M/H].

In Col. 'c' of Figs.~\ref{fig:fig2} to \ref{fig:fig6}, we keep the metallicity constant (at solar value), for two ages of 2 Gyr (blue line) and 12 Gyr (red line), respectively, and vary the slope of the bimodal IMF from 0.3 to 3.5 (x-axis). A general decrease of all CO features towards bottom-heavy IMFs can be seen.  \ion{Ca}{i}1.95, \ion{Ca}{i}1.99a and \ion{Ca}{i}1.99b indices change by $\sim$1 \AA \space between an SSP of $\Gamma_{b} = 0.3$ and $\Gamma_{b} = 3.5$, compared to $\sim$0.4 \AA \space for \ion{Ca}{i}1.99c and \ion{Ca}{i}2.26. Note that most of the NIR indices can be used to determine the IMF slope of populations with a steeper IMF than the Kroupa-like one, without any degeneracy between IMF and age/metallicity. For instance, the line-strength of CO1.64, in Col. 'c' of Fig.~\ref{fig:fig4}, changes from $\sim$3 \AA \space to $\sim$2.2 \AA \space for an old population when IMF slope varies from 1.3 to 3.5. According to Col. 'b', the minimum index value that can be derived by varying total metallicity from +0.26 to -0.35 dex, is $\sim$2.8 \AA. This means that index values less than $\sim$2.8 \AA \space can be only obtained for a bottom-heavy distribution. However, one should always consider how indices respond to  abundance ratios; for many NIR indices (e.g. FeH0.99, \ion{Na}{i}1.14, \ion{Mg}{i}1.18, \ion{Mg}{i}1.50, \ion{Na}{i}2.21, \ion{Fe}{i}2.24) the change due to elemental abundances (Col. 'd') is comparable to IMF variations, i.e. there is a (well-known) IMF-abundance ratio degeneracy.

Some indices are strongly sensitive to a given parameter. For example, \pab \space is much more sensitive to the age than metallicity or IMF. Both NIR sodium indices, \ion{Al}{i}1.31, \ion{Mg}{i}1.50, calcium indices in the \textit{K}-band and most of CO indices are strongly sensitive to IMF. Some indices are also singificantly sensitive to more than one parameter. For instance, CN indices are sensitive to age, but also to metallicity and IMF. 

Since the line strength indices, which we have discussed here, show different sensitivities to age, metallicity, IMF slope and elemental abundances, in principle it should be possible to break degeneracies between these parameters by employing an appropriate combination of indices. However, as we will see in Sect.~\ref{sec:example_applications}, this task is hampered by current limitations of SSP models in the NIR spectral range.

\subsection{Abundance ratio effects}\label{sec:abundance}

Panels in Col. 'd' of Figs.~\ref{fig:fig2} to ~\ref{fig:fig6}, plot the effect of element-by-element abundance changes on spectral indices, based on CvD12 SSP models. The changes are represented by vertical lines that start from a fiducial model of 13.5 Gyr, solar abundance and Chabrier IMF (dotted horizontal line), reaching index values for models with given elemental enhancement, as labeled at bottom of Col. 'd'. Notice that the lines show relative changes in index value as E-MILES and CvD12 models differ in absolute values for many line-strengths.  According to CvD12 models, \ion{Ca}{i}1.95 and \ion{Ca}{i}1.99b show a strong dependence on [C/Fe]. Fe indices are expected to have some sensitivity to the abundance of $\alpha$-elements. All CN indices response strongly to variations of C, N and $\alpha$. The variation in $\alpha$ for CN indices mostly comes from oxygen. The exception is CN1.46 which does not change with [O/Fe]; rather it varies with [Mg/Fe]. The Pa$\beta$ index at 1.28 \micron \space shows a mild dependency on titanium abundance. This is due to contamination of the central bandpass with \ion{Ti}{i} lines (see panel 'c' of Fig.~\ref{fig:figA1}). All CO indices have a strong dependence on carbon abundance, but CO1.60 and CO1.68 show a mild dependence on [Mg/Fe] too.

\subsection{Index dependence on the velocity dispersion and resolution}\label{sec:velocity_dispersion}

Since galaxies have intrinsic velocity broadening due to their velocity dispersion, it is important to test how line-strengths are affected by line broadening. This is also important to take into account the effect of instrumental resolution. Ideally, one would like to have indicators that are insensitive to variations in $\sigma$, such as those due to uncertainties in velocity dispersion measurements. The dependence of line-strength indices on velocity dispersion is quantified in Col. 'e' of Figs.~\ref{fig:fig2} to ~\ref{fig:fig6}. To this effect,  we  have convolved base E-MILES SSPs  to  different  velocity  dispersions in steps of 50\kms \space from 100 to 400\kms. The four fiducial models that we used have an age of 12 Gyr, two different metallicities and two different bimodal IMF slopes. The trends for SSPs with bimodal IMF of slope 1.3 are shown with solid lines, for solar (green) and super-solar (black) metallicities, respectively. For a bottom-heavy IMF with slope 3.0, the same trends are shown with dashed lines. The panels show that the dependence on $\sigma$ of broad molecular indices, such as CO and CN indices, is (as expected) barely noticeable. The line-strength of \textit{K}-band indices shows a significant change with $\sigma$, but for most of the indices, for $\sigma < 200$\kms \space (low- and intermediate-mass galaxy regime) the change in index-strength is modest. FeH0.99 is practically insensitive to velocity dispersion up to 400\kms. Almost all indices do not change their sensitivity to broadening with metallicity and IMF slope variations (comparing green and black colours, dashed and solid lines, respectively). Notice that, in most cases, the index-strength decreases with $\sigma$ above 200\kms.

\subsection{Wavelength shifts and uncertainties}\label{sec:wavelength_shifts}

Wavelength shifts due to galaxy internal rotational velocity, recessional velocity or uncertainties in wavelength calibration, can lead to variation in the measured index value. To evaluate these effects, we shifted a representative E-MILES SSP spectrum of age 12 Gyr, [M/H] = 0.06 and $\Gamma_{b}=1.3$ to radial velocities of $\pm$ 200\kms \space in steps of 4\kms. The index measurements as a function of radial velocity are shown in Col. 'f' of Figs.~\ref{fig:fig2} to ~\ref{fig:fig6}. According to these plots, some indices such as FeH0.99 in \textit{J}-, \ion{Mg}{i}1.50 in \textit{H}- and \ion{Na}{i}2.21 in \textit{K}-band require a rather careful wavelength calibration as they are very sensitive to radial velocity uncertainties (the required accuracy for those sensitive indices is provided in the plots). The same applies also to, e.g., the CO1.56 index in the \textit{H}-band. In fact, an uncertainty greater than 48 \kms in radial velocity  implies variations in the CO1.56 index comparable to those for a varying IMF, hence making the index not useful for an IMF analysis.

\subsection{Signal-to-noise ratio effects}\label{SNR}

\begin{landscape}
\begin{figure}
	\includegraphics[width=\linewidth]{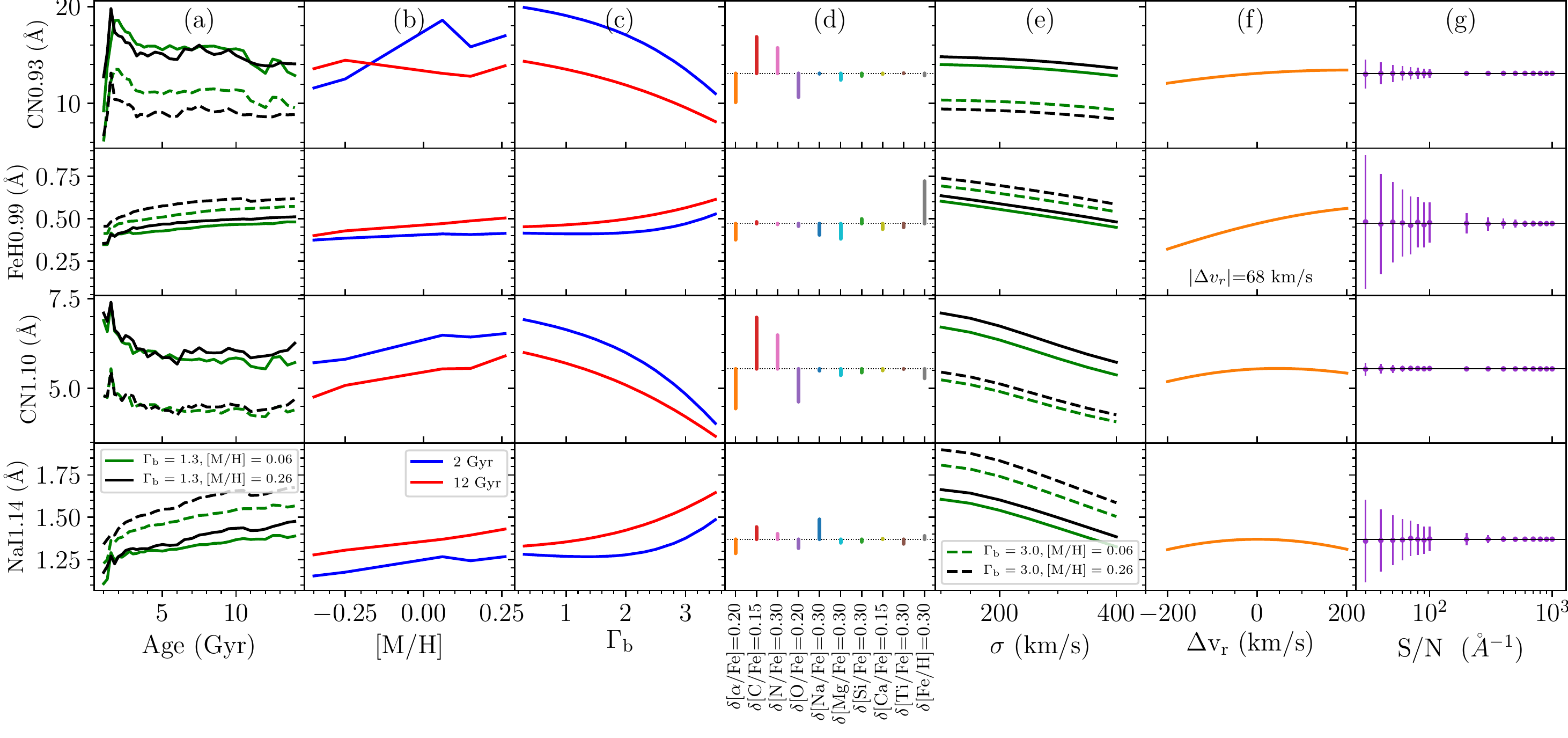}
    \caption{Model predictions of line-strength indices, defined in this work, as a function of age (panels in Col. 'a'), metallicity  (Col. 'b'), IMF slope (Col. 'c'), elemental abundance ratios (Col. 'd'), velocity dispersion (Col. 'e'), radial velocity (Col. 'f') and SNR (Col. 'g'). All plots are based on E-MILES models with BaSTI evolutionary tracks, but those in Col. 'd', where we used CvD12 SSPs to estimate the effect of elemental abundance variations. (Panels 'a') Different colors correspond to different metallicity, i.e. [M/H] = +0.06 (green) and [M/H] = +0.26 (black). Models with a Kroupa-like IMF are shown as solid lines while models with a bottom-heavy IMF ($\Gamma_{b}=3.0$) are shown as dashed lines. (Panels 'b') The blue lines are model predictions for young populations with an age 2 Gyr while the red lines correspond to old (12 Gyr) populations. (Panels 'c') Blue and red lines are predictions for young (2 Gyr) and old (12 Gyr) populations, respectively. (Panels 'd') The vertical lines show the variation of a given line-strength for variations of different elemental abundances, [X/Fe]'s, shown with different colours. (Panels 'e') Different colors correspond to different metallicities, i.e. [M/H] = +0.06 (green) and [M/H] = +0.26 (black), respectively. Models with a Kroupa-like IMF are shown as solid lines while models with a bottom-heavy IMF ($\Gamma_{b}=3.0$) are shown as dashed lines. (Panels 'f') The orange line shows index measurements on a reference E-MILES SSP spectrum shifted to a given radial velocity, $\rm \Delta v_r$. (Panels 'g') The error bars show the average uncertainties on index values  as a function of S/N.
    }
    \label{fig:fig2}
\end{figure}
\end{landscape}

\begin{landscape}
\begin{figure}
	\includegraphics[width=\linewidth]{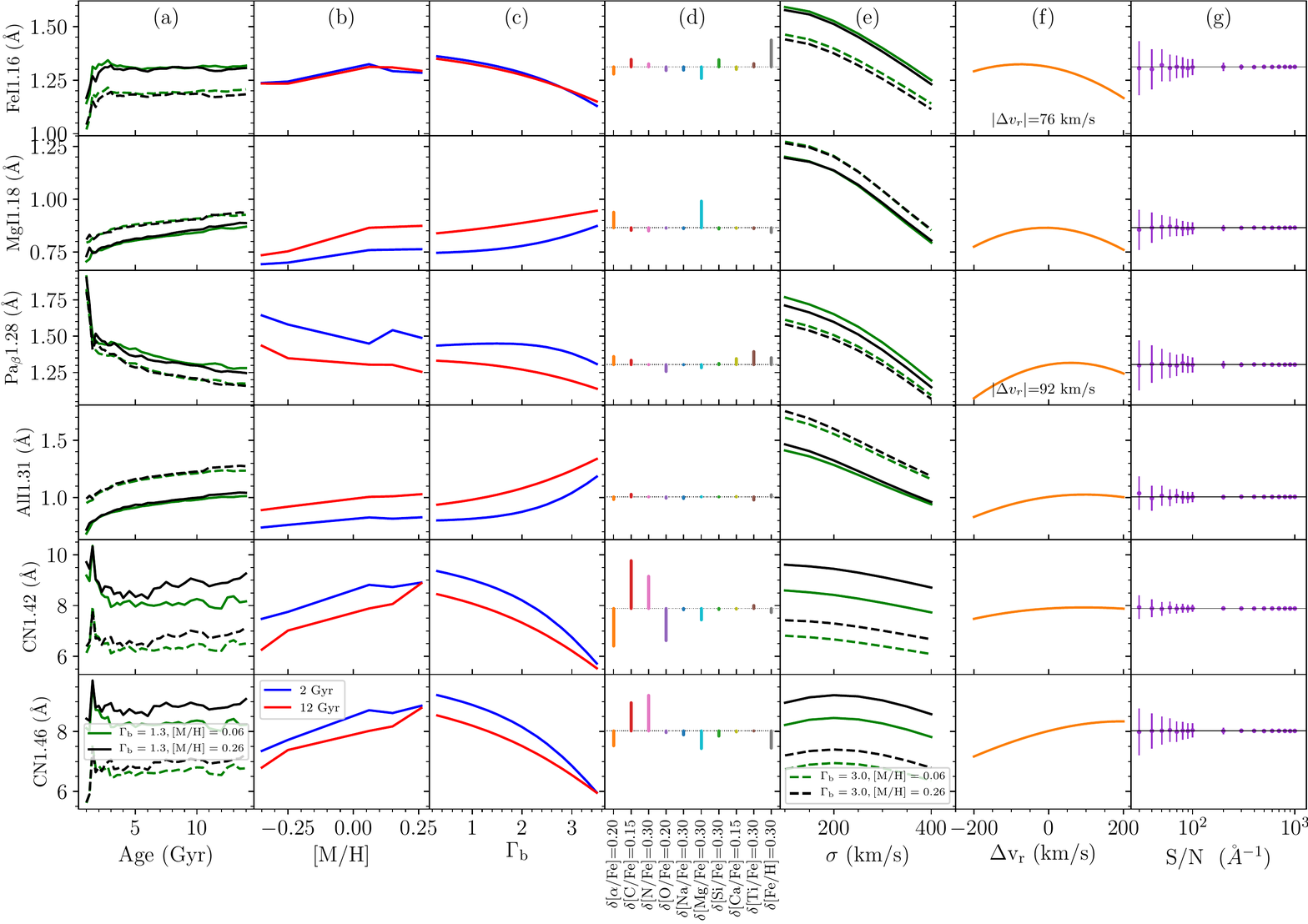}
    \caption{same as Fig.~\ref{fig:fig2}}
    \label{fig:fig3}
\end{figure}
\end{landscape}

\begin{landscape}
\begin{figure}
	\includegraphics[width=\linewidth]{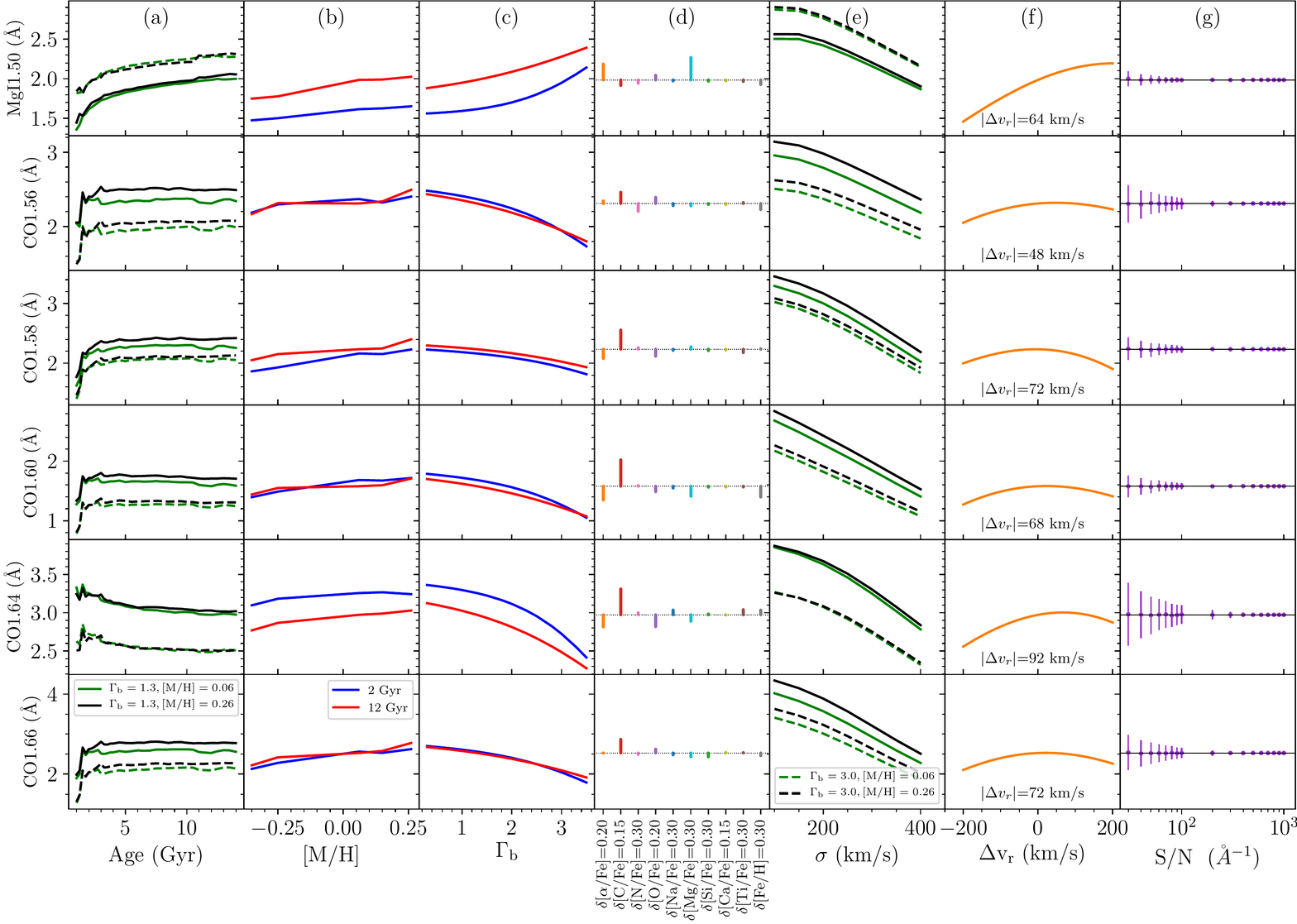}
    \caption{same as Fig.~\ref{fig:fig2}}
    \label{fig:fig4}
\end{figure}
\end{landscape}

\begin{landscape}
\begin{figure}
	\includegraphics[width=\linewidth]{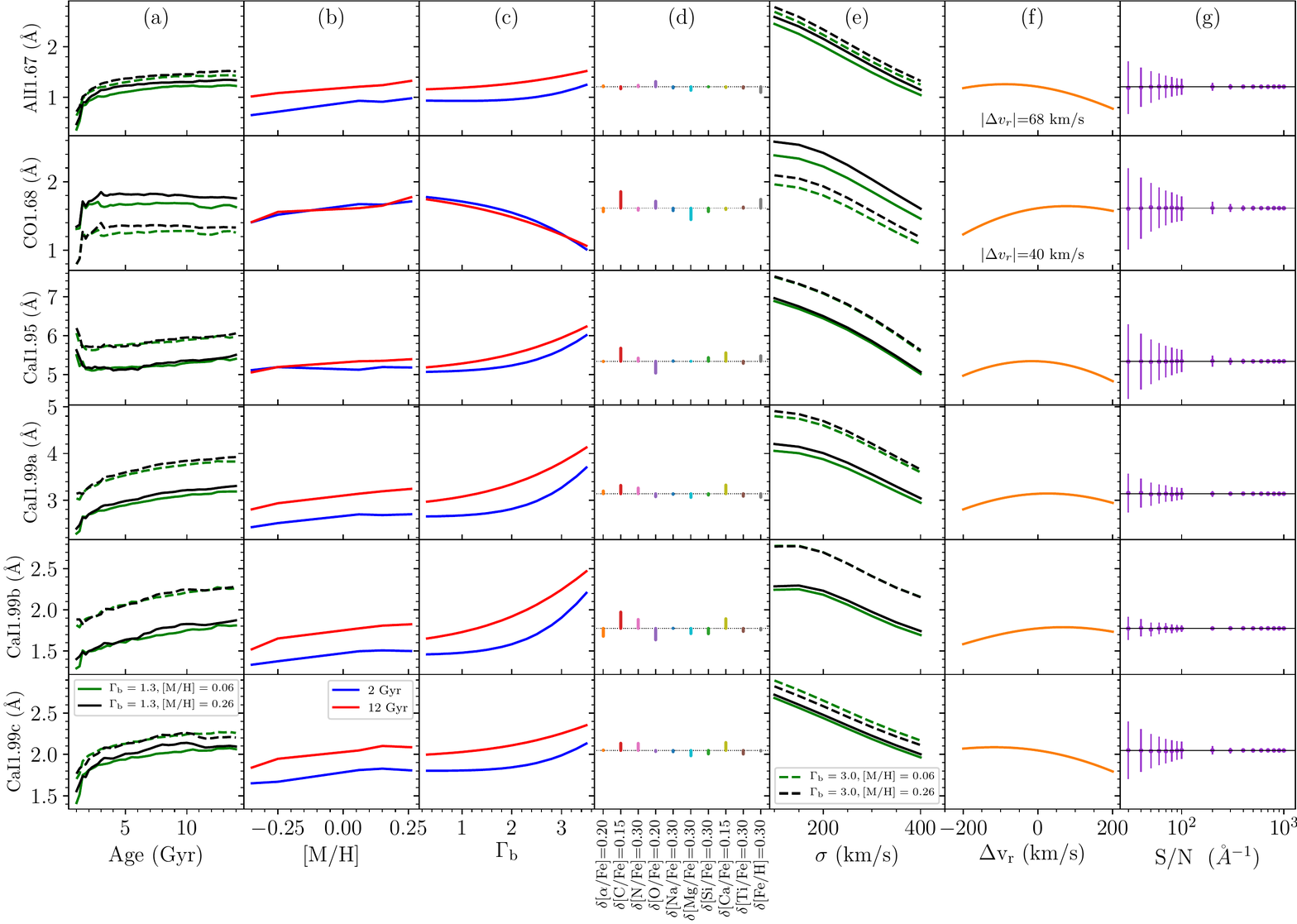}
    \caption{same as Fig.~\ref{fig:fig2}}
    \label{fig:fig5}
\end{figure}
\end{landscape}

\begin{landscape}
\begin{figure}
	\includegraphics[width=\linewidth]{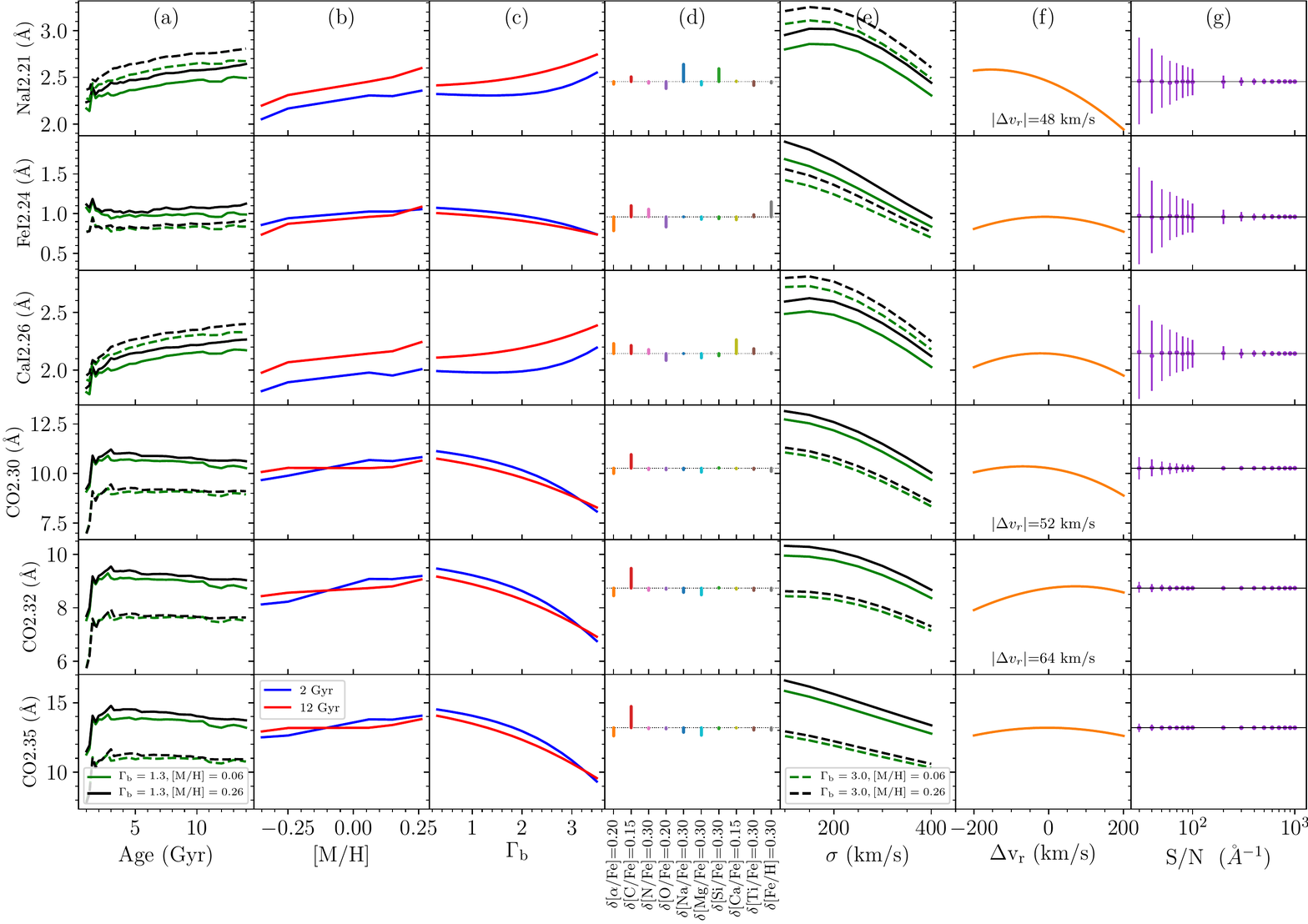}
    \caption{same as Fig.~\ref{fig:fig2}}
    \label{fig:fig6}
\end{figure}
\end{landscape}

Figures~\ref{fig:fig2} to ~\ref{fig:fig6} also show the sensitivity of spectral indices to SNR. The error bars in Col. 'g' show the average uncertainties on index measurements for a simulated set of E-MILES spectra with SNRs  in  the  range  of  30  to  1000 \AA$^{-1}$. The reference spectrum (horizontal black line) used to simulate the effect of different SNRs has an age of 12 Gyr, solar metallicity and bimodal IMF of slope 1.3. The error bars can be used to establish the required SNR to differentiate between two values of a given parameter with a given spectral index. For instance, in the case of FeH0.99 (Fig.~\ref{fig:fig2}), the SNR per angstrom required to differentiate between a Kroupa-like IMF and $\Gamma_{b}$ = 3.5 IMF is 200 \AA$^{-1}$.  For most of the indices, a SNR $\sim$ 100 \AA$^{-1}$ is needed to differentiate between a Kroupa-like IMF and a bottom-heavy IMF. In general, for indices covering a wider wavelength range, the required SNR is less demanding. For instance, for CO2.32 and CO2.35, a spectrum with SNR of 30 \AA$^{-1}$ would be sufficient to distinguish between a metal-poor population ([M/H]=-0.25) and a metal-rich one ([M/H]=+0.15).

\begin{table}
	\centering
	\caption{Impact of flux calibration on line-strength indices. Columns 1 and 4 give the name of indices sorted by increasing width. The width of each index ($\rm \lambda_{red2} - \lambda_{blue1}$) is provided in Cols. 2 and 5. The fractional change on indices value when removing continuum is given in Cols. 3 and 6.}
	\label{tab:tab2}
	\begin{tabular}{c c c c c c} 
		\hline\hline
		Index	& Width	& $\rm \frac{\Delta I}{I_{o}}$	& Index	& Width	& $\rm \frac{\Delta I}{I_{o}}$\\
		  	& (\AA)	& (\%) &   	& (\AA)	& (\%)	\\
		(1) & (2)	& (3) & (4) & (5) & (6)	\\
		\hline

Pab1.28  & 115.0  & 0.2 & CO1.66  & 235.0  & 2.1 \\
FeH0.99  & 150.0  & 7.4 & AlI1.67  & 235.0  & 2.9 \\
AlI1.31  & 150.0  & 0.3 & NaI2.21  & 255.0  & 0.0 \\
CO1.58  & 155.0  & 0.4 & CaI2.26  & 265.0  & 0.4 \\
CO1.68  & 165.0  & 1.4 & CO2.30  & 320.0  & 17.5 \\
NaI1.14  & 175.0  & 0.4 & CN1.10  & 325.0  & 7.2 \\
CO1.64  & 175.0  & 0.1 & CO2.32  & 335.0  & 4.1 \\
MgI1.50  & 185.0  & 0.4 & CN1.42  & 335.0  & 11.5 \\
CO1.60  & 200.0  & 0.3 & CO2.35  & 370.0  & 27.1 \\
FeI1.16  & 200.0  & 0.5 & CN1.46  & 380.0  & 4.1 \\
CO1.56  & 225.0  & 0.7 & CaI1.99b  & 430.0  & 5.8 \\
FeI2.24  & 230.0  & 0.3 & CaI1.99a  & 430.0  & 2.9 \\
MgI1.18  & 230.0  & 0.2 & CaI1.99c  & 430.0  & 5.6 \\
CaI1.95  & 230.0  & 0.9 & CN0.93  & 490.0  & 4.8 \\

		\hline
	\end{tabular}
\end{table}


\subsection{Index sensitivity to flux calibration}\label{flux_calibration}

Following \citet{vazdekis1999b}, we studied potential systematic effects on line-strengths due to flux calibration by first measuring indices on an E-MILES SSP spectrum (i.e. flux-calibrated) and then on a continuum subtracted one. In Cols. 3 and 6 of Table~\ref{tab:tab2}, we show, for each index, the fractional change (in percent) due to continuum subtraction. The width of each index (from the bluest to the reddest end of the pseudo-continua) is shown in Cols. 2 and 5 of the table. Indices are sorted by increasing width to illustrate a general trend of increasing fractional change with increasing index width. However, we notice that the trend is not linear. For instance, the fractional change of \ion{Ca}{i}1.99a which is 430 \AA \space wide is similar to that of CO1.66 and \ion{Al}{i}1.67 whose definitions span less than 300 \AA. Indices involving wavelength ranges narrower than 300 \AA \space have negligible (mostly less than 1\%) fractional changes, except for FeH0.99. \newline

\subsection{Observed wavelengths of indices as a function of redshift}\label{sec:redshift}

\begin{landscape}
\begin{figure}
	\includegraphics[width=\linewidth]{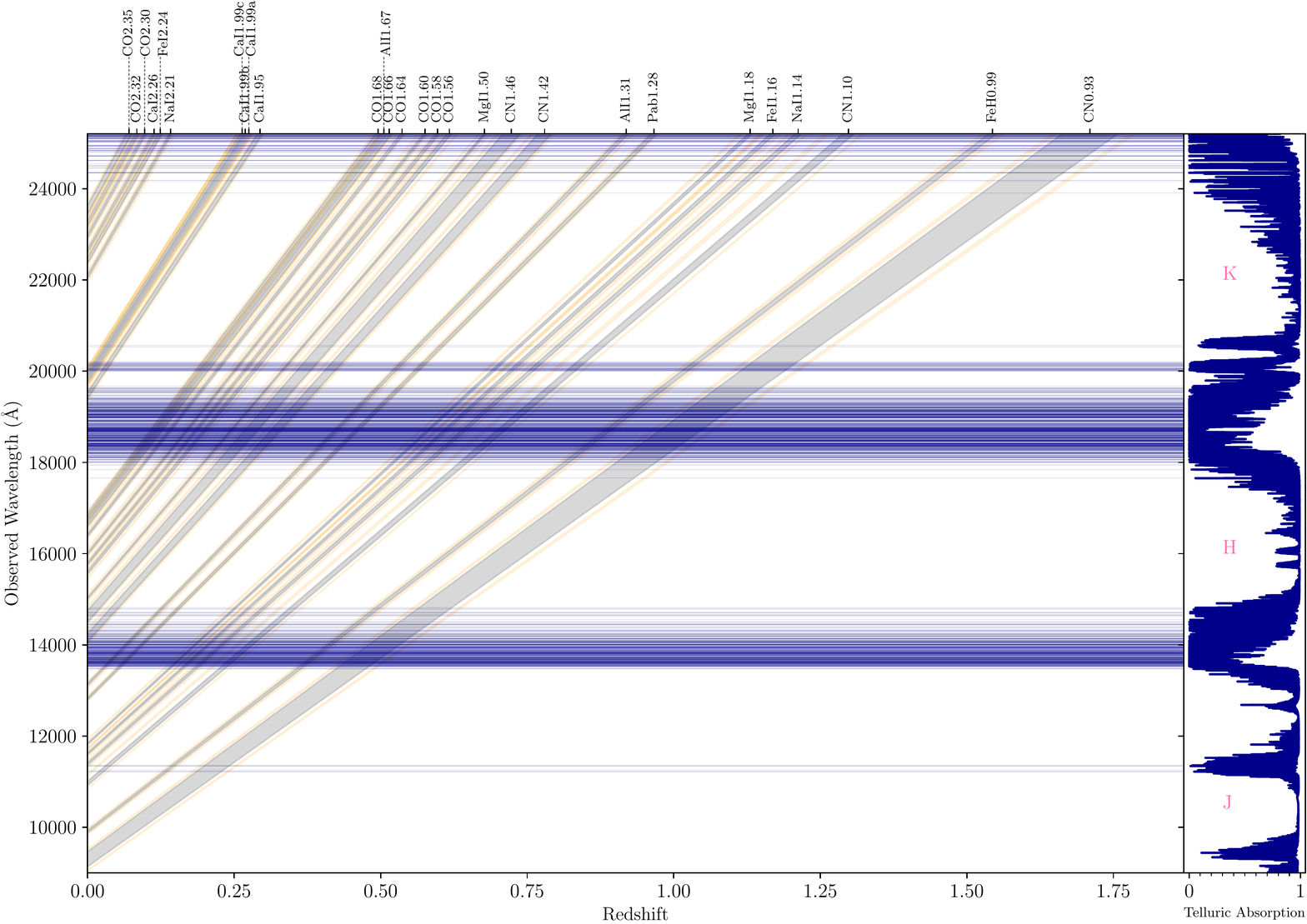}
    \caption{Observed bandpasses of spectral indices as a function of redshift. The grey and orange shaded areas correspond to the central and pseudo-continua bandpasses of each index, respectively. The right panel shows a telluric absorption spectrum, with strongest absorption lines (transmission of less than 10\%) being overplotted by horizontal dark blue lines in the left panel.}
    \label{fig:fig7}
\end{figure}
\end{landscape}

The wavelengths of spectral indices in Table~\ref{tab:tab1} are given in the rest-frame; with increasing redshift, the indices shift to redder wavelengths and might fall in regions with prominent telluric absorption features. Figure~\ref{fig:fig7} shows the observed wavelengths, defining each index bandpasses, as a function of redshift. The grey shaded area corresponds to indices' central bandpass, while the orange ones correspond to the pseudo-continua bandpasses. The indices are indicated on top of the figure. In the right panel, we show atmospheric transmission and indicate the strongest absorptions (transparency less than 10\%) with horizontal dark blue lines in the left panel. This plot should help the user to identify the best redshift ranges for studying specific spectral indices, or identify potentially useful spectral indices in a given observed-frame window. For instance, redshift ranges of $0.3<z<0.6$ and $0.8<z<1.25$, are not suitable for studying CN0.93 and FeH0.99 indices, as in these redshift ranges, both indices are severely contaminated by atmospheric absorption. As another example, if a spectrograph has a spectral coverage from 9000 to 14000 \AA, according to Fig.~\ref{fig:fig7}, \pab, and \ion{Al}{i}1.31 can be observed up to $z\sim0.05$, CN1.10, \ion{Na}{i}1.14, \ion{Fe}{i}1.16, and \ion{Mg}{i}1.18 up to $z\sim0.15$, while CN0.93 and FeH0.99  up to $z\sim0.5$

This figure also shows that it is possible to perform a stellar population analysis including at least three of the newly defined indices (i.e. CN0.93, FeH0.99, and CN1.10) out to $z\sim1.2$. Moreover, none of the NIR indices will fall within the J and H bands for $z > 1$.

\subsection{Reliability of measured indices in the NIR}\label{sec:reliability}

For defining wavelength limits of NIR indices, we tried to avoid as much as possible rest-frame regions affected by sky contamination. This ensures that when measuring indices on stellar population models, one should have clean estimates. However, depending on the redshift of galaxies, their rest-frame indices can be actually affected by sky emission and telluric lines. This can lead to deriving unreliable stellar population parameters. To avoid systematic effects due to sky contamination or other potential sources of uncertainty (e.g. bad pixels and cosmic rays), we provide a general prescription to assess the robustness of each index analysis.

\begin{figure}
	\includegraphics[width=\columnwidth]{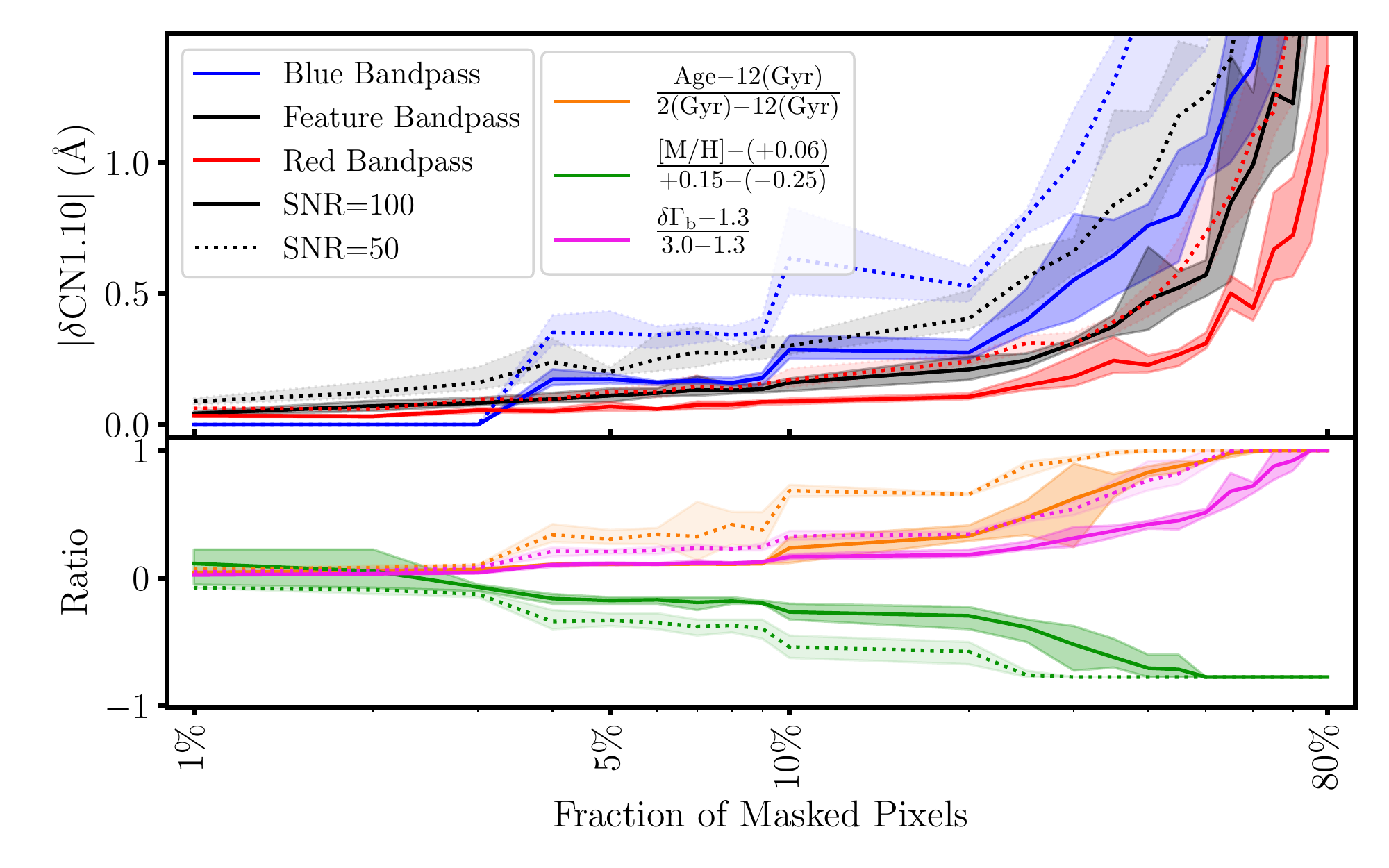}
    \caption{Effect of removing a given percentage of pixels within blue (blue solid (dotted) line), feature (black solid (dotted) line) and red (red solid (dotted) line) bandpasses on the line-strength of CN1.10 index for a spectrum with SNR=100 (50) \AA$^{-1}$ (upper panel). The lower panel shows the variation of stellar population paramaters that correspond to the change of index values due to the masking of pixels within the index bandpasses. The changes in stellar population parameters are computed with respect to a reference spectrum with age 12 Gyr, [M/H]=0.06, and $\Gamma_{b}=1.3$, and are normalized to range of stellar populations parameters of intermediate-mass and massive ETGs (see the text). Shaded areas, in both the upper and lower panels, show the effect of varying the velocity dispersion of the spectra from 60 to 360 \kms. For a spectrum with SNR=100 \AA$^{-1}$, assuming a maximum relative variation of $\pm$0.5 for stellar population parameters, the lower panel shows that the fraction of masked pixels has to remain lower than $25\%$, $25\%$, and $45\%$, to obtain acceptable estimates of age, metallicity, and IMF, respectively. For a smaller SNR this fraction is smaller as described in the text.}
    
    \label{fig:fig8}
\end{figure}

To this aim, we tried to establish the maximum percentage of affected pixels in a spectral range of an index for which a reliable index estimate is still possible. In practice, one may consider as affected pixels those for which the residual of any kind of correction is just at, or above, the noise level. To address the problem, we used an E-MILES SSP as our reference spectrum and removed a given percent of pixels randomly (from $1\%$ to $80\%$) within either the blue pseudo-continuum, or feature, or red pseudo-continuum bandpasses, respectively. The reason for doing so is that, in principle, whether the affected pixels reside in the pseudo-continua or the feature, it affects the line-strength differently. For each index, the process was repeated 1000 times, and the index line-strength was measured for each iteration. To be conservative, for each bandpass, we saved the maximum change in the line-strength with respect to that of the reference spectrum. This maximum variation might correspond, for instance, to the case where adjacent pixels in a given bandpass have to be masked out because of, e.g., the poor subtraction of blended sky lines. We performed this procedure on spectra with various resolutions from 60 to 360 \kms. Moreover, we considered the effect of SNR by using two sets of spectra with signal-to-noise ratios of 100 and 50 per Angstrom, respectively. Figure~\ref{fig:fig8} shows the results for the CN1.10 index as an example. Solid lines in the upper panel show the changes in the index measurement as a function of the fraction of removed pixels within blue pseudo-continuum bandpass (blue), feature bandpass (black) and red pseudo-continuum bandpass (red) with shaded areas including the effect of varying velocity dispersion  for spectra with SNR=100 \AA$^{-1}$, while the dotted lines show the same but for spectra with SNR=50 \AA$^{-1}$. We adopt as reference model an E-MILES SSP with an age of 12 Gyr, solar metallicity and Kroupa-like IMF. In the lower panel of the Figure, we have transformed the changes of index value (taking the average among bandpasses) to variations in stellar population parameters. Hence, the lower panel of Fig.~\ref{fig:fig8} shows the effect of masking a given fraction of pixels on the derived stellar population parameters. The orange lines show the variation of age with respect to the reference spectrum, normalized to an age range of 10 Gyr (between 2 to 12 Gyr). The variation of metallicity with respect to the reference spectrum is shown with green lines. Metallicity variations are normalized to a range from [M/H] = -0.25 to [M/H] = +0.15 (i.e. 0.4~dex). The pink lines correspond to the changes of bimodal IMF slope with respect to the reference spectrum, and it is normalized to an IMF slope range from 1.3 to 3.0 (i.e. 1.7). As for the upper panel, solid and dotted lines show results for spectra with SNR=100 \AA$^{-1}$ and SNR=50 \AA$^{-1}$, respectively, while shaded regions correspond to variations of stellar population parameters with velocity dispersion, from 60 to 360 \kms. Notice that most intermediate-mass and massive ETGs fall within the ranges of parameters that we considered for the above normalization (see, e.g., \citealt{labarbera2013}). By considering a cut-off limit of half the total sensitivity to the relevant stellar population parameters (ratio=$\pm$0.5), this figure shows that one can robustly measure the age, metallicity and IMF slope (on a spectrum with SNR=100 \AA$^{-1}$), if the number of affected pixels within the CN1.10 bandpasses does not exceed $25\%$, $25\%$, and $45\%$ of the total pixels in each bandpass limit, respectively. For spectra with lower SNR, the maximum fraction of affected pixels within bandpasses is smaller. For example, for the CN1.10 index, when constraining age, metallicity and IMF, a SNR of 50 \AA$^{-1}$ implies a maximum fraction of affected pixels within the bandpasses of $9\%$, $9\%$, and $25\%$, respectively. We notice that, while we considered here the case of a cut-off limit of $\pm0.5$, in general, one should decide the cut-off threshold based on the specific application and the data quality.

Appendix~\ref{sec:appendixB} shows the same plot as Fig.~\ref{fig:fig8} but for all indices defined in this paper for both SNR of 100 and 50 \AA$^{-1}$. Table ~\ref{tab:tabB1} in Appendix~\ref{sec:appendixB} provides, for each index, the maximum number of affected pixels for which the index can be still used  for meaningful stellar population analysis. We recommend that, when using spectral indices for stellar population studies, one should identify contaminated pixels in the data and see whether the fraction for the index of interest is less than those provided in Table~\ref{tab:tabB1}. If this is the case, one should measure the index by removing (more precisely, by interpolating) the contaminated pixels and be confident that the index is still useful for the study, otherwise the index should be discarded. A practical example has also been provided in the appendix, which explains in detail how a user can take benefit of the numbers given in the table. Note that this table is intended to provide a quick reference for users, while Figs.~\ref{fig:figB1}, ~\ref{fig:figB2}, and ~\ref{fig:figB3} allow for a more flexible criterium selection, depending on the desired application.

We point out that the benefit of the approach outlined above is that it can be performed on any sample of galaxies at any redshift. Also, it is relatively general in such a way that it takes into account not only the effect of pixels contaminated because of uncertainties due to sky subtraction or telluric correction, but also because of any other issue deriving from data reduction. Moreover, the same approach could also be implemented to characterize other sets of indices, e.g. in the  UV and optical spectral ranges, as when studying high redshift galaxies, the UV/optical features are redshifted to the NIR and thus are unavoidably plagued by sky contaminations. Overall, we suggest using our approach as a standard procedure in the analysis of (NIR) spectral features.  

\section{Example Applications}\label{sec:example_applications}

We discuss here some applications of our newly defined indices to the study of unresolved stellar populations. To this effect, we rely on E-MILES and CvD12 spectral synthesis models, selecting samples of ETGs from the literature. The selected samples cover a range of morphological types (from ellipticals to spirals), velocity dispersions (from $\sim$30 to $\sim$360\kms) and environments (field and cluster).

\subsection{Galaxy samples}\label{sec:galaxy_samples}
\begin{itemize}

\item\citet{labarbera2019}: The sample consists of seven nearby massive ETGs ($\sigma_{0} > 300$\kms) selected from SPIDER survey \citep{labarbera2010} and SDSS DR7 \citep{abazajian2009}. Most of these galaxies are brightest cluster galaxies, while two of them are satellites (XSG2 and XSG7). The high-quality spectra were obtained with the X-Shooter multi-wavelength medium resolution slit echelle spectrograph \citep{vernet2011} at the ESO Very Large Telescope (VLT). The NIR part of the spectra covers a wavelength range from 9800 to 25000 \AA \space with a resolution power of $\sim$5500, which allows an accurate correction of sky lines. All galaxy spectra have SNR above 100 \AA$^{-1}$. For our study, we used galaxy spectra for the innermost radial bins defined in \citet{labarbera2019}, as their quality and SNR is highest. In addition to the spectra of individual galaxies, we also used a stacked spectrum for all seven galaxies. The typical age, metallicity, IMF slope, and $\alpha$-enhancement of the central regions of galaxies in this sample is 11 Gyr, +0.26 dex, 3.0, and +0.4, respectively. 

\item\citet{francois2019}: The sample consists of 14 bright nearby galaxies ranging from ellipticals to spirals along the Hubble sequence. The authors observed the galaxies from the optical to the NIR (3000-24800 \AA), using the X-Shooter spectrograph \citep{guinouard2006} at the ESO-VLT on Paranal (Chile). They obtained the spectra with a resolution power between R $\approx 4000$ and $R \approx 5400$ depending on the wavelength and slit width and sampled the same spatial region in arcsecond of galaxies to minimize any systematic effect originating from stellar population gradients. This sample spans a wide range in velocity dispersion (from 36 to 335\kms, according to table 3 in \citet{francois2019}).

\item\citet{baldwin2018}: Twelve ETGs of this sample are drawn from the $\rm ATLAS^{3D}$ survey \citep{cappellari2011} spanning a narrow range of velocity dispersion from ~80 to ~120 \kms. This sample covers a wide range of star formation histories as previous works in the literature show that some of the galaxies in this sample have regions of ongoing star formation. NIR spectra were obtained with the Gemini Near-Infrared Spectrograph on the Gemini north telescope in Hawaii. The spectra cover the wavelength range 8000-25000 \AA \space at a resolution $R \sim$ 1700. 

\item\citet{silva2008}: The sample includes six elliptical galaxies and three S0 galaxies in the Fornax cluster. Long-slit spectra of the central regions ($\rm < \frac{1}{8}R_{eff} $) of these galaxies were obtained with the ISAAC NIR imaging spectrometer at the ESO-VLT. The wavelength coverage of the spectra is $\sim 21200-23700$ \AA \space with a resolving power of R $\approx 2900$. The velocity dispersion of this sample spans a range between 128 to 222\kms.

\item\citet{marmol2008}: The sample consists of twelve elliptical and S0 field galaxies and their spectra were obtained with the ISAAC NIR imaging spectrometer at the ESO-VLT. The instrumental configuration was the same as in \citet{silva2008} in order to allow a direct comparison of the two samples. The lowest velocity dispersion in this sample is 59\kms, while the highest one is 305\kms. \newline

\end{itemize}

We convolved the spectra of \citet{labarbera2019}'s galaxy sample to $\sigma=360$\kms, as this is an upper limit to the individual velocity dispersions of galaxies in different samples. The reduced spectra of galaxies in the \citet{francois2019} and \citet{baldwin2018} samples are publicly available. While the spectra of \citet{francois2019} are rest-frame, we have used the Heliocenteric velocities provided in Table 1 of \citet{baldwin2018} to shift their galaxy spectra to the rest-frame. Both samples are flux calibrated, and we only needed to convolve their spectra out to  $\sigma=360$\kms. The reduced spectra of \citet{silva2008} and \citet{marmol2008} samples are the same as those used by \citet{rock2017}, who corrected them to restframe, and convolved to a common resolution of $\sigma=360$ \kms. Note that the spectra in these samples are extracted within $\rm < \frac{1}{8}R_{eff} $ except for galaxies in \citet{francois2019} which are extracted within $\rm < 0.3R_{eff} $. As all spectra correspond to galaxy central regions, we do not expect a significant aperture bias in our derived results.

\subsection{Constraining the shape of the low-mass end of the IMF}\label{sec:application1}

Assuming a given shape for the IMF affects the derivation of several observable properties of galaxies, such as mass-to-light ratios and star formation rates. Therefore, the IMF plays a fundamental role in understanding galaxy evolution mechanisms. The difficulty in measuring the low-mass end of the IMF is that variations in the number of low-mass stars (which dominate the mass budget of ETGs)  imply only little changes in the integrated light of old stellar populations. According to \citet{labarbera2013}, although IMF-sensitive indices in the optical can constrain the mass fraction of low-mass stars expected in the IMF, they cannot be used to distinguish the functional form of the IMF. \citet{conroy2012a} proposed that the shape of the low-mass IMF in old stellar populations can be constrained by a combination of multiple IMF-sensitive indices, taking advantage of giant-sensitive features, such as  TiO0.89. The work by \citet{labarbera2016} has been the first attempt to combine TiO0.89 and the optical TiO1 and TiO2 features to separate the effect of the IMF from giant stars and abundance ratios. They found that the response of FeH absorption at 9900 \AA \space to a low-mass tapered (bimodal) IMF is much smaller with respect to that for a unimodal (single power-law) one, helping up to constrain the IMF functional form at the very low-mass end. 

\begin{figure}
    \centering
	\includegraphics[width=\columnwidth]{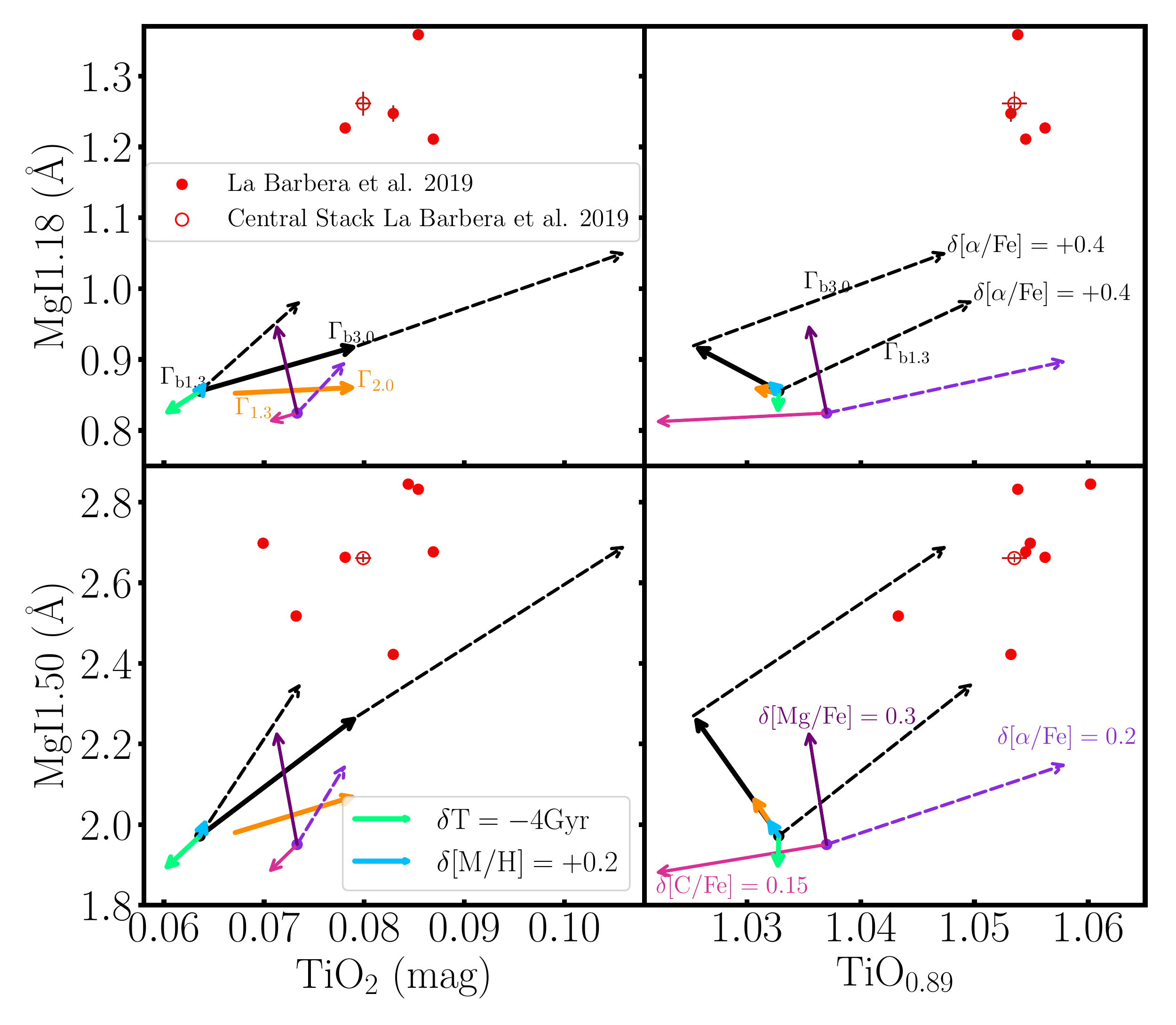}
    \caption{ Response of selected TiO and \ion{Mg}{i} spectral indices to variations in the IMF, abundance ratios and other stellar population parameters. Solid black (orange) arrows show the effect of varying the IMF from a Kroupa-like (Salpeter) to a bottom-heavy IMF, for E-MILLES SSP models, with an age of 11 Gyr, solar metallicity and $\rm[\alpha/Fe]=0.0$. The changes in \ion{Mg}{i} indices due to a varying unimodal IMF slope is very small compared to the bimodal IMF slope variation, allowing us, in principle, to constrain the shape of the IMF low-mass end. We also show the effect of a change in age (by -4 Gyr) and metallicity (by +0.2 dex) for the E-MILES model, for a Kroupa-like IMF population (green and blue arrows, respectively). Both \ion{Mg}{i} features have no dependence on [M/H], while they decrease very little with decreasing age. The effect of increasing \afe, in both E-MILES models with Kroupa-like IMF and bottom-heavy one, is shown by the dashed black arrows, while the purple and pink arrows correspond to variations of +0.3 and +0.15 dex in [Mg/Fe] and [C/Fe] from CvD12 models of age 13.5 Gyr and [Fe/H]=0.0. The red points show the \citet{labarbera2019} galaxy sample. Red open circles are the measurements for the stacked spectrum (see the text). All model and observation line-strengths refer to a velocity dispersion of 360\kms. Notice that while \ion{Mg}{i} and TiO2 indices are measured as equivalent widths, the TiO0.89 index is measured as a flux ratio.}
    \label{fig:fig9}
\end{figure}

As a possible application of the NIR spectral indices defined in the present work, we have tried to explore which indices can provide us with further relevant constraints on the IMF functional form. We have found that Mg indices might indeed help us to constrain the detailed shape of the IMF at intermediate masses. 
While performing a detailed analysis of this point is beyond the scope of the present work, we illustrate the overall idea in Figure~\ref{fig:fig9}, plotting two NIR \ion{Mg}{i} indices (MgI1.18 and MgI1.50) vs. two optical TiO indices (TiO2 and TiO0.89), measured for different stellar population models, namely E-MILES and CvD12. Since the high quality galaxy spectra from \citet{labarbera2019} span both optical and NIR wavelengths, and have been shown to exhibit a bottom-heavy IMF ($\Gamma_{b} > 2.6$, in the innermost radial bins), we only used this sample in the plots. The measurement of indices on individual galaxies is shown with red points while the open red circle corresponds to the stacked spectrum. Solid orange and black arrows show the effect of a change in the slope of unimodal and bimodal IMFs on the E-MILES models from $\Gamma = 1.3$ to $\Gamma = 2.0$ and $\Gamma_{b} = 1.3$ to $\Gamma_{b} = 3.0$, respectively. Dashed black arrows show the effect of enhancement of +0.4 dex in the abundance of $\alpha$ (typical value of the sample) for both a Kroupa-like IMF and a bottom-heavy one. Light blue and yellow arrows display the effect of a change in total metallicity and age of populations with Kroupa-like IMF. They change from solar metallicity to [M/H] = +0.26 dex (typical metallicity of the sample) and from 11 Gyr (mean age of the sample) to 7 Gyr, respectively. CvD12 models have been used in these plots as well. The solid purple arrow shows the effect of a change of +0.3 dex in magnesium enhancement while the dashed violet arrow corresponds to the $\alpha$-enhancement of +0.2 dex. As TiO0.89 is very sensitive to the abundance of carbon, we also show the effect of enhancing carbon by +0.15 dex with a pink arrow.

While the strength of \ion{Mg}{i} indices changes significantly by going from a Kroupa-like IMF to the bimodal IMF of slope 3 ($\rm \frac{\Delta I}{I_{o}}$ = 7.5\% for \ion{Mg}{i}1.18 and 15.1\% for \ion{Mg}{i}1.50), it almost does not change between a Salpeter IMF and a unimodal IMF of slope $\Gamma = 2$ \footnote{According to figure 12 of \citet{labarbera2013}, a unimodal IMF of slope 2 can describe the data equally well as a bimodal IMF of slope 3; both slopes provide similar $F_{\rm<0.5M_{\odot}}$ (M dwarfs fraction) values.} ($\rm \frac{\Delta I}{I_{o}}$ = 0.9\% for \ion{Mg}{i}1.18 and 4.7\% for \ion{Mg}{i}1.50). The differences due to different IMF shapes are subtle but measurable. Regarding the sensitivity of \ion{Mg}{i} features to other stellar population properties, E-MILES models show that both \ion{Mg}{i}1.18 and \ion{Mg}{i}1.50 are independent of total metallicity (see the light blue arrow). According to CvD12 models, both \ion{Mg}{i} features are sensitive to \afe \space and [Mg/Fe] (dashed violet and solid purple arrows) and \ion{Mg}{i}1.50 is sensitive to [C/Fe] (solid pink arrows) as well. The dashed black arrows show the effect of $\alpha$-enhancement on both a population with Kroupa-like IMF and a population with bottom-heavy IMF. These are obtained from $\alpha$-enhanced E-MILES models (an updated version of Na-enhanced models of \citet{labarbera2017}). The effect of $\alpha$-enhancement and bimodal IMF can be singled out in the plots of both \ion{Mg}{i}1.18 and \ion{Mg}{i}1.50 vs TiO0.89, as the dashed black and solid black arrows are almost orthogonal. Note that the age of E-MILES models is 11 Gyr which corresponds to the mean age of the \citet{labarbera2019}  sample while the age of CvD12 models is 13.5 Gyr. All indices are measured in EW except the TiO0.89 index, which is measured as a flux ratio. Unfortunately, the comparison of data and model predictions shows that none of the current models in Fig.~\ref{fig:fig9} is able to fully match the data. However, a bimodal IMF might partly explain the high value of \ion{Mg}{i} indices. 

\begin{figure}
\centering
	\includegraphics[width=.8\columnwidth]{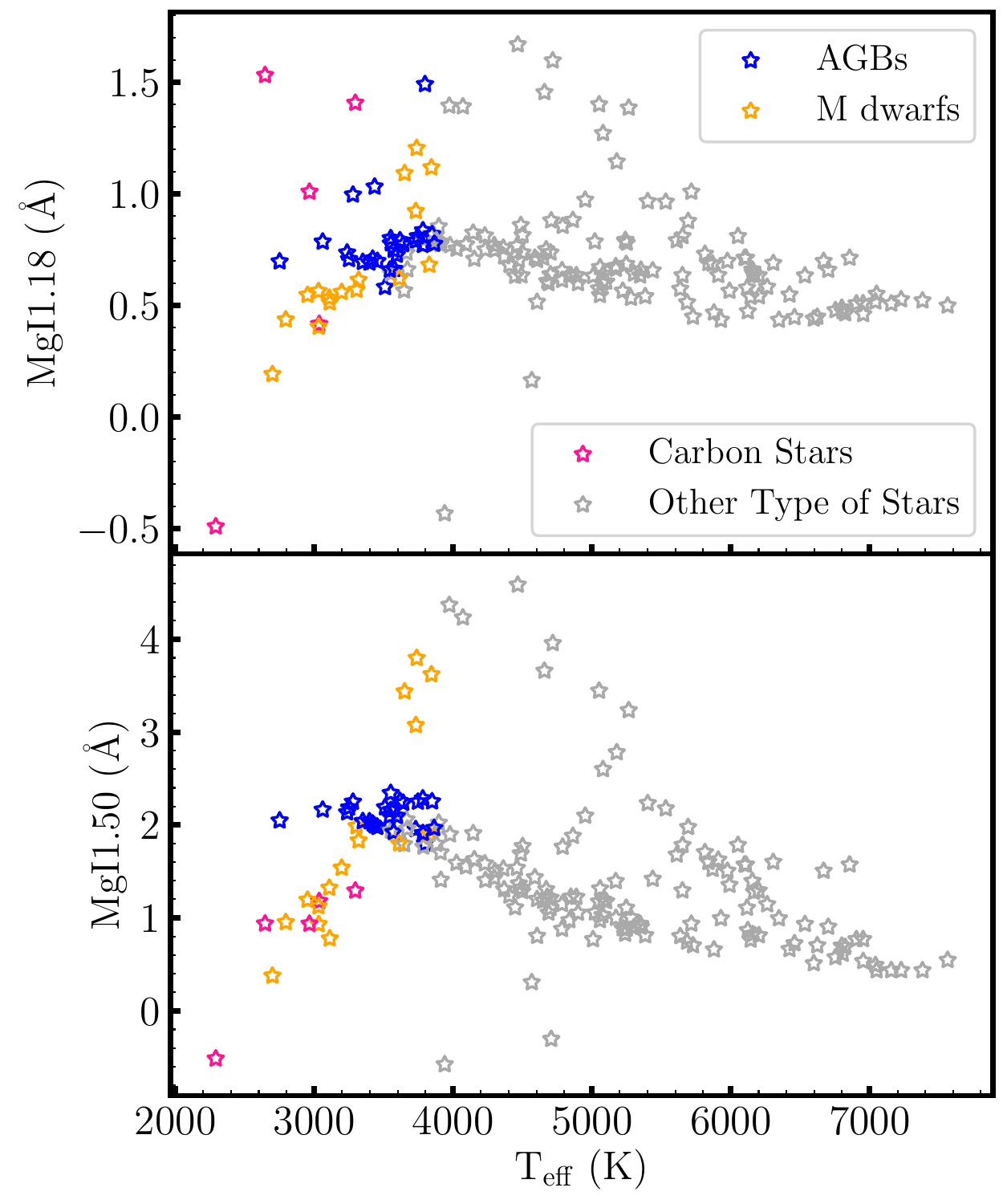}
    \caption{ Dependence of NIR \ion{Mg}{i} spectral indices on effective temperature for IRTF stars that have been used to compute E-MILES SSP models. Blue, orange, and pink colours display AGB, M-dwarf, and carbon stars, respectively. Other types of stars are represented with grey colour. In both plots, M-dwarf stars peak in strength at 3800K. 
}
    \label{fig:fig10}
\end{figure}

To understand why NIR \ion{Mg}{i} indices can distinguish between different IMF parametrisations we have looked at the IRTF stars used to construct the models. Figure~\ref{fig:fig10} shows the magnesium indices measurements on the spectra of IRTF stars as a function of their effective temperature. The type of stars (for stars cooler than 3900K) is from table 2 of \citet{rock2015}. M-dwarf stars (orange points) are the ones that correspond to the low-mass end of the IMF. These plots show that M-dwarf stars peak in strength around 3800K. According to figure 5 of \citet{vazdekis2012} the mass of these stars is around 0.4\ms. While bimodal IMF gets flatten for stars with masses less than 0.6\ms, the unimodal IMF, keep increasing to lower masses. This means that for a unimodal IMF, the dwarf stars at $\sim$3800K (which have the highest values of \ion{Mg}{i}) count less than the cooler ones (which have the lowest value of \ion{Mg}{i}). Therefore, by increasing the slope of the unimodal IMF, the overall value of \ion{Mg}{i} does not vary significantly. However, by increasing the slope of the bimodal IMF, the number of stars with masses less than 0.6\ms, all increase with similar weight,  and stars with the highest value of \ion{Mg}{i} end up to dominate the overall absorption of \ion{Mg}{i}. 

\subsection{On the need of further developments for NIR SPS modelling}\label{sec:application2}

We find a significant discrepancy when comparing some NIR line-strength indices from SSP models and data for ETGs. This shows that current SPS models have difficulties at reproducing galaxy spectra in the NIR (e.g. \citet{riffel2019}).

\begin{figure*}
	\includegraphics[width=\linewidth]{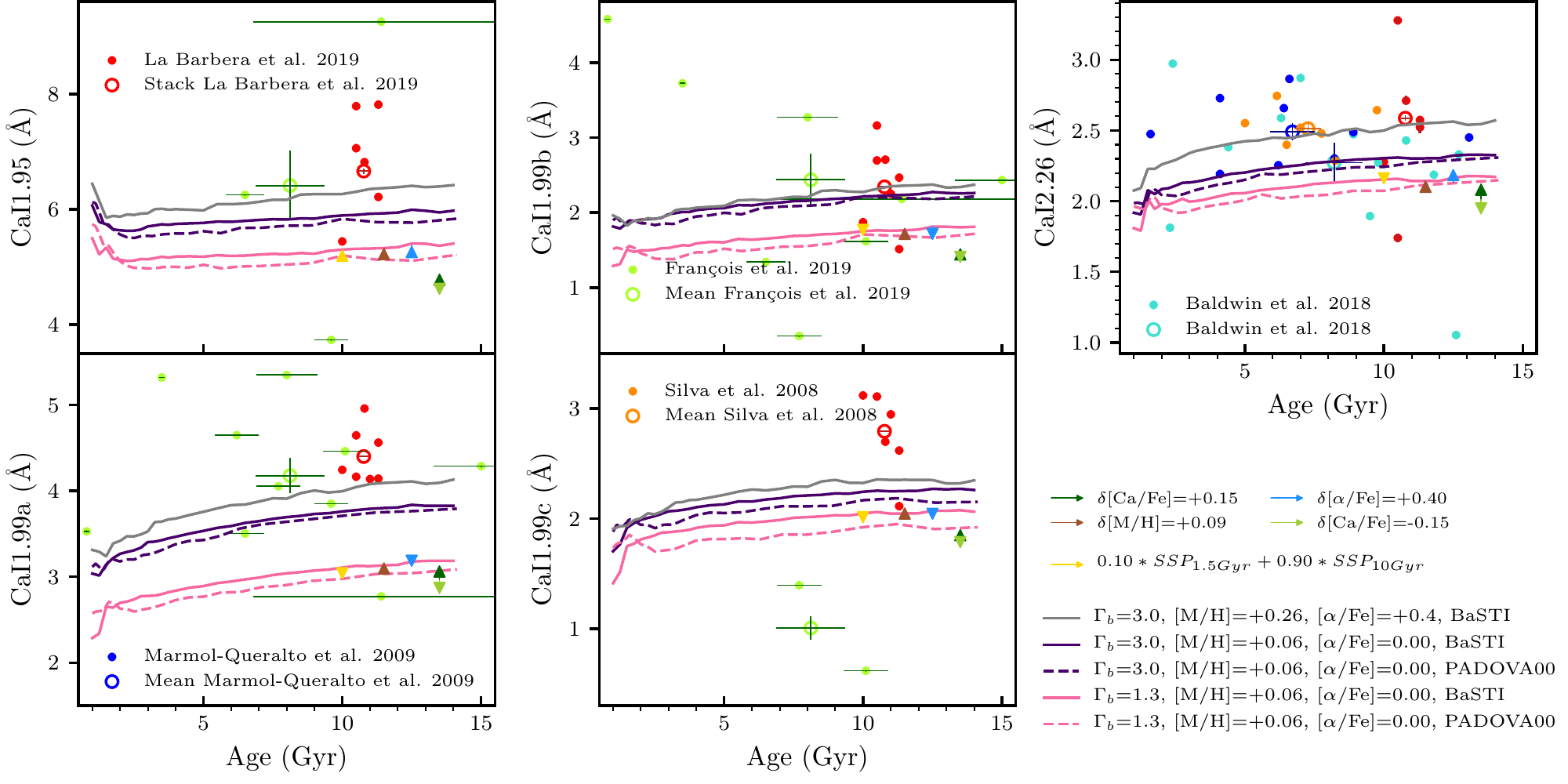}
    \caption{ \ion{Ca}{i} line-strength indices in the \textit{K}-band as a function of age. The solid purple and pink lines display the predictions from scaled-solar E-MILES (BaSTI-based) models for two different IMF slopes, while the dashed lines correspond to the models with the same parameters but based on PADOVA00 isochrones. Overplotted are the measurements of the same index for the ETGs of \citet{silva2008} (orange points), \citet{marmol2008} (blue points),  \citet{baldwin2018} (cyan points), \citet{francois2019} (green points) and for the massive ETGS of \citet{labarbera2019} together with their respective mean values including errors (open circles). The thick grey line is representative of the population of \citet{labarbera2019} sample. The yellow arrow at 10 Gyr marks the change in the \ion{Ca}{i} indices when adding a mass fraction of 10\% of the young population of 1.5 Gyr to the underlying SSP model of 10 Gyr. The brown arrow, at 11.5 Gyr, indicates the change in the \ion{Ca}{i} indices when moving from solar metallicity model to a population with a total metallicity of 0.15. The light blue arrow at 12.5 Gyr denotes the effect of $\alpha$-enhancement of +0.4 dex on a scaled-solar population. The effect of an increase/decrease in calcium abundance from CvD12 models is shown with dark/light green arrows at 13.5 Gyr.}
    \label{fig:fig11}
\end{figure*}

As a further example, in Fig.~\ref{fig:fig11}, we show \ion{Ca}{i} indices in the \textit{K}-band measured from various samples of galaxies and compare them with predictions of E-MILES SSPs as a function of age. The red points correspond to the central regions of massive ETGs from \citet{labarbera2019}, while the red open circle shows the corresponding stacked spectrum. Line-strengths for galaxies from \citet{francois2019} are also shown by yellow green points, together with their mean values, plotted as yellow green open circles. Orange, blue,  and cyan points show measurements for \citet{silva2008}, \citet{marmol2008}, and \citet{baldwin2018} galaxies while orange, blue, and cyan open circles correspond to their mean values. The pink and violet solid lines display the predictions from base E-MILES models with solar metallicity for two different IMF slopes as obtained with BaSTI isochrones, whereas the dashed lines illustrate the respective line-strengths measured from models obtained with PADOVA00 isochrones. As our samples cover a wide range in velocity dispersion, we have used SSPs with $\Gamma_{b}=1.3$ and $\Gamma_{b}=3$ (pink and violet) to account for the expected variation of IMF slope with velocity dispersion.  The thick grey line corresponds to predictions of E-MILES SSPs with IMF slope of 3.0, $\alpha$-enhancement of 0.4 dex, and total metallicity of +0.26 dex, i.e. typical values of stellar population  properties for the \citet{labarbera2019} sample. In the case of 'perfect' stellar population models, one would expect the grey line to pass through the open red circles in all panels of Fig.~\ref{fig:fig10}.

For all calcium features in the \citet{labarbera2019} sample, the predictions from E-MILES models can only reproduce the galaxies with the smallest \ion{Ca}{i} values, but most of the galaxies show larger line-strengths (notice, however, the large dispersion in data). The \ion{Ca}{i}1.99b and \ion{Ca}{i}2.26 indices of the \citet{labarbera2019} stacked spectrum are reasonably well-matched by models with bottom-heavy IMF ($\rm \Gamma_b=3.0$), $\alpha$-enhancement of 0.4 dex, and total metallicity of +0.26 dex. We point out that $\alpha$-enhanced E-MILES models are safe at such high metallicity as for constructing the models with [$\alpha$/Fe] = +0.4, one relies on IRTF stars with [Fe/H] = -0.04 (according to equation 4 of \citet{vazdekis2015}). However, the models (grey lines) do not fit the other three indices, although they approach the data in the right direction. For what concerns the data of \citet{francois2019}, in all panels the sample shows a huge scatter. Such scatter can be due to either a real dispersion among galaxies, or (more likely) to some contamination of the indices from sky residuals. In \ion{Ca}{i}2.26 plot, none of the galaxies from \citet{francois2019} sample are shown as for all galaxies, this index is flagged as unreliable according to the procedure described in Sect.~\ref{sec:reliability}. In general, the \ion{Ca}{i}2.26 shows significant scatter between 2.0 and 3.0 \AA. This scatter has been already reported by \citet{silva2008} who could not provide a clear explanation for it. They point out, however, that the \ion{Ca}{i} feature is hampered by the contribution of many other absorbers such as S, Si, Ti, Sc and Fe, although a significant contribution of these elements to the index strength  is not seen when using CvD12 models (see Col. 'd' of Fig.~\ref{fig:fig6}). 

The disagreement between the predicted and the observed line strength in the NIR has been attributed to an enhanced contribution of stars in the asymptotic giant branch (AGB) evolutionary phase (see the discussion in \citet{marmol2008}). These stars dominate the NIR light of intermediate-age ($\sim$ 1-2 Gyr) stellar populations. Therefore, we added a mass fraction of $10\%$ of a young population of age 1.5 Gyr to the underlying standard SSP model of 10 Gyr in order to study whether the predictions of such model might result into a better fit to the observed \ion{Ca}{i} indices. It is worth stating that, indeed, this mass fraction is higher than what expected from optical indices, such as, in particular, the Balmer lines \citep{labarbera2013}. The yellow arrow at 10 Gyr indicates the change in the \ion{Ca}{i} indices due to the additional contribution of the young component on top of the old one. The size of the arrow is tiny. As a result, such model is unable to fit the \ion{Ca}{i} indices of ETGs. Also the arrow points to the opposite direction than the required to match the data (with the exception of  \ion{Ca}{i}1.95). This suggests that the most likely contribution of evolved stars is actually the one from old stellar populations. 

The effect of increasing total metallicity from solar ([M/H] = +0.06) to [M/H] = +0.15 dex, in the E-MILES model is shown by the brown arrow at 11.5 Gyr, while the effect of increasing \afe \space by +0.40 dex is shown by a sky blue arrow at 12.5 Gyr. According to these arrows, an increase in metallicity does not increase the \ion{Ca}{i} index-strengths significantly. However, an $\alpha$-enhancement can improve the models only for the \ion{Ca}{i}2.26 index. It is worth noting that the NIR part of the E-MILES models is based on 180 stars of IRTF library which are mostly around solar metallicities and solar abundance ratios. Therefore predictions of models with super solar metallicity are less reliable (see \citet{rock2015}), and would require stellar libraries spanning a wider metallicity range.

The arrows at 13.5 Gyr, mark the change in the \ion{Ca}{i} indices when moving from a reference model to a CvD12 model of [Ca/Fe] = +0.15 (dark green) or [Ca/Fe] = -0.15 (light green). Note that there is an offset between predictions of the CvD12 models and E-MILES models at scaled-solar abundance with similar stellar population parameters. This could be due to differences in several modelling ingredients, such as selection of stars in (different) stellar libraries, isochrones, parameterization for the IMF and even different values for the solar metallicity \citep{vazdekis2015}. For the present purposes, we are more interested in the relative responses of CvD12 models to variations of [Ca/Fe]. Also, notice that CvD12 models with varying abundance ratios are computed at a fixed age of 13.5 Gyr by considering a Chabrier IMF. Also, the combined effect of elemental abundances and IMF variations is not fully understood yet, as discussed in the case of Na lines by \citet{rock2016} and \citet{labarbera2017}. In the latter, the authors developed a Na-enhanced version of E-MILES models and showed that the effect of [Na/Fe] enhancement is not linear, having a larger effect for some Na indices as the IMF slope becomes steeper. Also, while the effect of calcium abundance on the strength of calcium indices in the optical domain such as \ion{Ca}{ii}0.39, \ion{Ca}{i}0.42 and CaT is significant, according to these plots, this effect is minor/negligible on NIR \ion{Ca}{i} indices. Moreover, unlike in the optical range (see \citet{vazdekis1997}), the NIR \ion{Ca}{i} indices are above (not below) model predictions.

Comparing the effect of IMF slope on NIR calcium indices (pink and violet lines) with the effect of abundances (blue, dark and light green arrows), shows that these lines have larger sensitivity to IMF than abundance ratios at least as far as current models seem to predict. This was first noticed by \citet{conroy2012a}. Since the optical calcium indices are more sensitive to abundances than IMF, they suggested that a combination of calcium indices in the optical and NIR spectral ranges should be able to disentangle the degeneracy between IMF and abundance ratios.

Note also that by varying the IMF from Kroupa-like to bottom-heavy the models get significantly closer to all NIR calcium lines (as seen by comparing pink and purple lines in  Fig.~\ref{fig:fig11}), although not enough to properly match them.

The interesting point about the behaviour of calcium indices in the NIR is that they all behave in a similar way for what concerns the mismatch to observations. Hence, any future modelling of stellar populations should be able to behave consistently among all NIR calcium lines.

It is noteworthy to mention that the discrepancy between observations and models in plots of Fig.~\ref{fig:fig11} cannot be attributed to the index definition, as similar mismatch can be seen if other definitions are used for the same absorption. \citet{baldwin2018} considered (see their fig.~17) the line-strength of the calcium absorption at 2.26 $\mu$m following the index definition of \cite{frogel2001}. The authors compare their sample of low-mass ETGs with predictions of various models, including E-MILES. 
The reason why they did not find any significant discrepancy with resepct to E-MILES, is, indeed, the large scatter and error bars on observed line-strengths, making all models in the plot compatible with the data.

Fig.~\ref{fig:fig11} shows that the solution  from the optical line-strengths (grey line), matches the mean value of the observations (red open circles) for the \ion{Ca}{i}1.99b and \ion{Ca}{i}2.26 indices. However, for \ion{Ca}{i}1.95, \ion{Ca}{i}1.99a and \ion{Ca}{i}1.99c indices, the optical solution does not match the data but it helps to reduce the discrepancy between model predictions and observed line-strengths, in particular for \ion{Ca}{i}1.95 and \ion{Ca}{i}1.99a. Also note that for these two indices the mismatch between the best (optical) solution and the observations is significantly smaller
than the difference obtained when varying the IMF slope. In order to achieve a good match to observations, modelling of integrated stellar populations in the NIR will have to further improve, especially for what concerns the behaviour in the high metallicity regime as well as the effect of abundance ratios. Moreover, it is important to confront in a detailed manner the models to both NIR and optical indices, to provide fully consistent solutions over a large wavelength baseline. This point will be further explored in a forthcoming paper.

\section{Summary} \label{sec:summary}

In this work, we have defined a new system of spectral indices in the NIR which are optimized and fully characterized for studying stellar population of unresolved systems. 

With the aid of E-MILES stellar population models, we have obtained responses of SSP spectra to changes in age, metallicity and stellar IMF. Such responses, once associated with absorption lines, have guided us to identify features that can be used as indicators of relevant population parameters in the NIR. 

We defined new NIR indices in a similar manner to the Lick system, where spectral indices are defined with a central bandpass covering the spectral feature and two other bandpasses at the red and blue sides of the feature to trace the local continuum. We defined the central bandpass of each index to encompass the maximum extent of the peak of the sensitivities around the absorption feature itself and avoid, if possible, contamination by stellar population metallic lines, telluric absorptions or sky emissions. In this way, we optimised the indices to be sensitive to the main population parameters, namely age, metallicity and IMF. Concerning continuum bandpasses, we have defined the limits in wavelength regions weakly affected by absorption features and less contaminated by telluric absorptions or sky emissions (this applies to stars and galaxies at z=0). Furthermore, the spectral bandpasses were defined as wide as possible to make them less sensitive to degradation due to the velocity dispersion or spectral resolution and also to increase the SNR of the index measurements. However, the full wavelength coverage for most indices is shorter than $\sim$300 \AA, to minimise flux calibration issues.

Depending on the redshift of galaxies, their absorption features may fall in regions not clean from telluric absorption and sky emission lines. The pixels within the index definition that have not been corrected well during the data reduction can affect the spectral index values at a level comparable to that due to variations of stellar population parameters. In order to overcome this problem, we have outlined a prescription to identify indices that can be safely measured in a given stellar population study. In this method, pixels containing substantial systematic uncertainties should be masked out before performing index measurements. If the fraction of masked pixels within the index bandpasses is less than a maximum fraction (which can be adopted depending on the specific application), which is provided in Table~\ref{tab:tabB1}, the index can be safely used for the stellar population analysis (in particular to estimate age, [M/H], and IMF). If such constraints are not met, the index should be excluded altogether.

The present work provides the first comprehensive characteristics of NIR spectral indices. To this effect, we used base E-MILES models, as well as CvD12 models with variable abundance ratios. The analysis shows that most of the NIR indices require high SNR ($\sim$100) to obtain useful measurements. 

We illustrate the potential use of the newly defined set of NIR indices with two applications to the general population of ETGs. First, we show that it is possible to constrain the shape of the IMF at low-mass end, using a combination of magnesium indices in the NIR and TiO indices in the optical. The analysis also shows that it is possible to break the degeneracy between IMF and abundance ratios. In the second application, we confront several NIR Ca line-strengths of ETGs from the literature with predictions from SPS models. We find that predictions of models with super-solar metallicity, \afe \space enhanced abundance ratios, and bottom-heavy IMF provide reasonably good match to two, out of five, NIR calcium lines as measured from a stacked spectrum representative of the central regions of the most massive ETGs. On the other hand, the models underestimate the line-strengths for the remaining three calcium lines. Although the overall solution found from this analysis seems to agree with that derived from the optical range, the comparison highlights the need of improving significantly  model predictions in the super-solar metallicity and non-solar abundace ratio  regimes, in particular for what concerns  the contribution  of very cool stars, which dominate the NIR light.

The system of NIR indices presented in the present work should offer new venues to constrain the stellar population content of galaxies in the \textit{JWST} era. To this effect, the plots presented here could be taken as a guide for preparing and optimizing future observations to study stellar populations based on NIR indices.

\section*{Acknowledgements}
We thank Micheal Beasley for discussions and his help and support at different stages of this work. We are grateful to the anonymous referee for her/his helpful comments that allowed us to  significantly improve our manuscript. The authors acknowledge support from grants AYA2016-77237-C3-1-P and PID2019-107427GB-C32 from the Spanish Ministry of Science, Innovation and Universities (MCIU).  This work has also been supported through the IAC project TRACES which is partially supported through the state budget and the regional budget of the Consejer\'\i a de Econom\'\i a, Industria, Comercio y Conocimiento of the Canary Islands Autonomous Community. This paper made use of the IAC Supercomputing facility HTCondor (\url{http://research.cs.wisc.edu/htcondor}), partly financed by the Ministry of Economy and Competitiveness with FEDER funds, code IACA13-3E-2493.

\section*{Data Availability}

The E-MILES SSP models are publicly available at the MILES website (\url{http://miles.iac.es}). The updated version of Na-enhanced models of \citet{labarbera2017} are also available from the same website (under "Other predictions/data"). The \citet{conroy2012a} SSP models are available upon request to the authors (see \url{https://scholar.harvard.edu/cconroy/projects}). Observations of \citet{labarbera2019} sample made with ESO Telescope at the Paranal Observatory under programmes ID 092.B-0378, 094.B-0747, 097.B-0229 (PI: FLB). The central spectra and stacked central spectrum are available from FLB upon request. The spectra of \citet{baldwin2018} sample were taken using the Gemini Near-Infrared Spectrograph on the Gemini North telescope in Hawaii through observing program GN-2012A-Q-22. The reduced spectra (FITS files) are available via \url{https://github.com/cbaldwin1/Reduced-GNIRS-Spectra}. Observations of \citet{francois2019} sample made with ESO Telescopes at the La Silla Paranal Observatory under programme ID 086.B-0900(A). The reduced spectra (FITS files) are available via \url{http://cdsarc.u-strasbg.fr/viz-bin/qcat?J/A+A/621/A60}. Observations of \citet{silva2008} sample performed at the European Southern Observatory, Cerro Paranal, Chile; ESO program 68.B-0674A and 70.B-0669A. Observations of \citet{marmol2008} sample performed at the European Southern Observatory, Cerro Paranal, Chile, as well. \citet{rock2017} corrected the data of these two samples to restframe and convolved them to a resolution of $\sigma=360$ \kms and they are available from EE upon request. The IRTF Spectral Library is observed with the SpeX spectrograph, at the NASA Infrared Telescope Facility on Mauna Kea and the spectra are publicly available at \url{http://irtfweb.ifa.hawaii.edu/~spex/IRTF_Spectral_Library/}.



\bibliographystyle{mnras}
\bibliography{references} 




\appendix

\section{Spectral Windows of the New Indices and Other Index Definitions from the Literature} \label{sec:appendixA}

In this section, we zoom into a region around each index and display atomic and molecular contributors to the index. As for some of the indices, there are multiple definitions in the literature, we compare the limits of the bandpasses in the definitions for NIR indices and discuss their trends, as well.

\begin{figure*}
\centering
	\includegraphics[width=.88\linewidth]{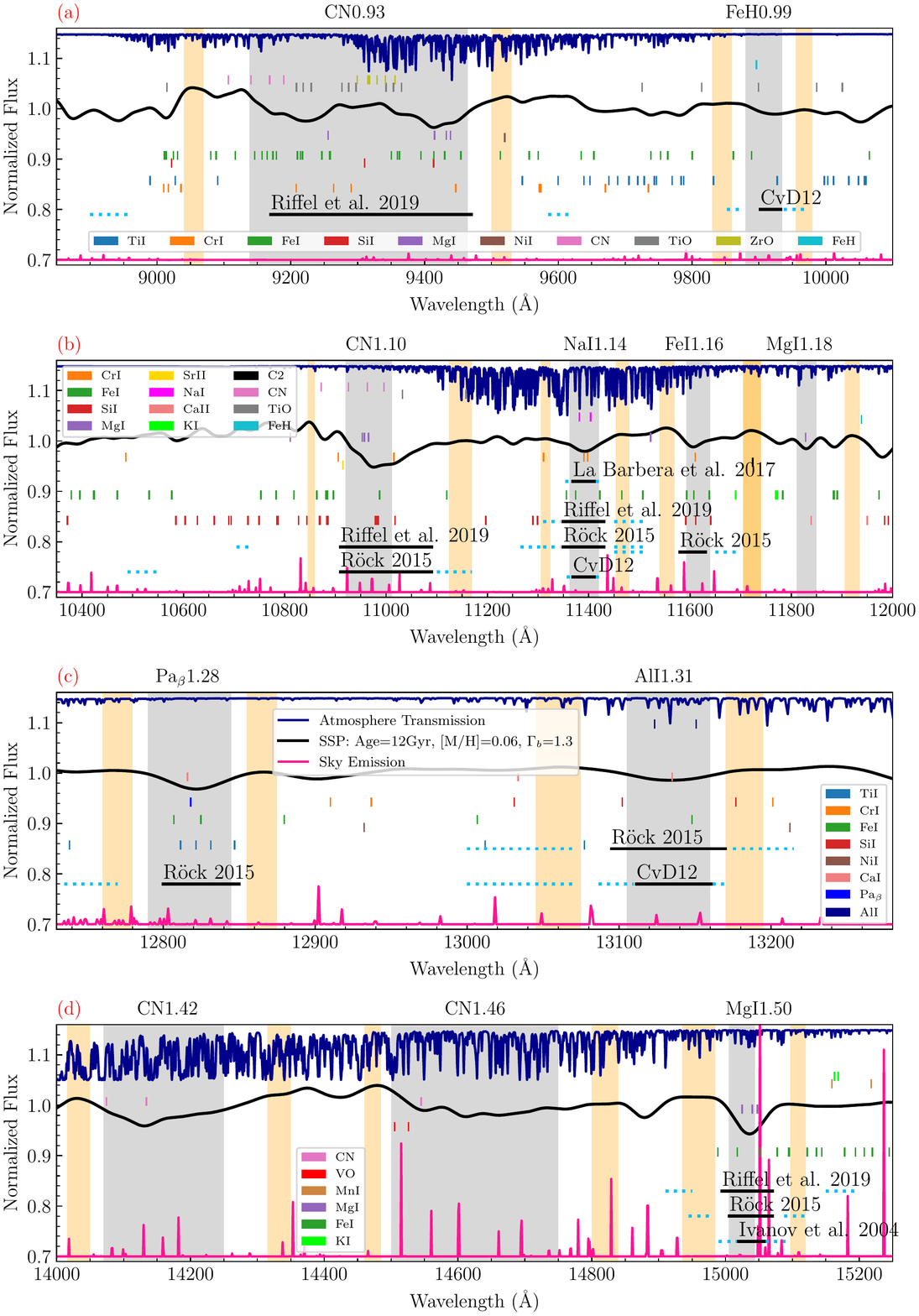}
    \caption{Zoom into the region of NIR indices defined in this work and comparison of the various definitions in the literature. The thick black line, in each panel, shows a normalised E-MILES SSP spectrum of age 12 Gyr, [M/H] = +0.06 and bimodal IMF slope of 1.3, broadened to velocity dispersion of 360\kms. The dark blue and pink lines correspond to the atmospheric transmission and sky emission spectra from ESO Skycalc tool. The small vertical lines show the position of atomic and molecular absorptions from \citet{kleinmann1986}, IRTF library and \citet{lagattuta2017}. Grey and orange regions represent central absorption and pseudo-continua bandpasses in our definition while the corresponding bandpasses in the definitions from the literature is shown by solid black and dashed blue horizontal lines.}
    \label{fig:figA1}
\end{figure*}

\begin{figure*}
\centering
	\includegraphics[width=.95\linewidth]{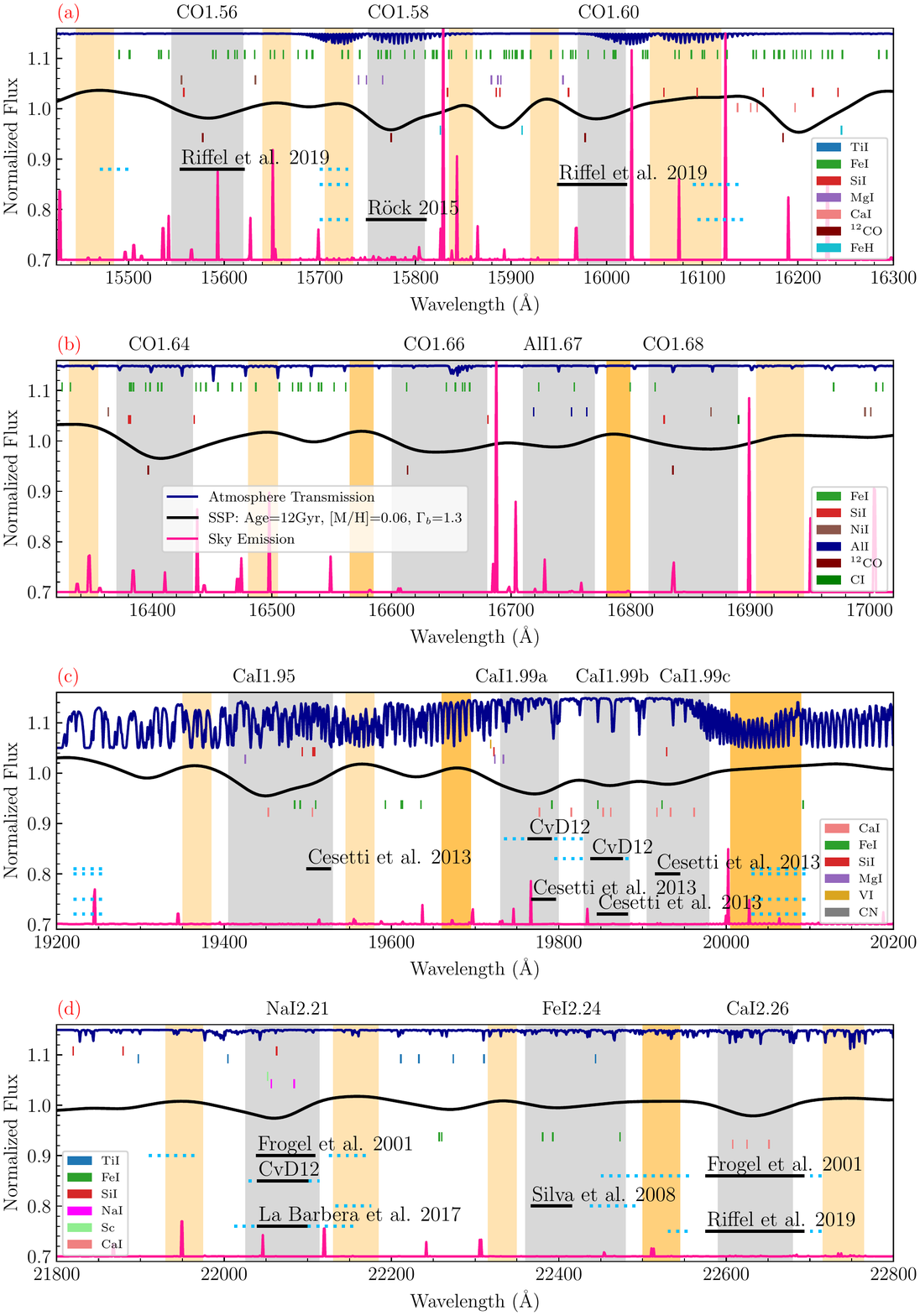}
    \caption{Same as Figure~(\ref{fig:figA1})}
    \label{fig:figA2}
\end{figure*}

\begin{figure*}
\centering
	\includegraphics[width=\linewidth]{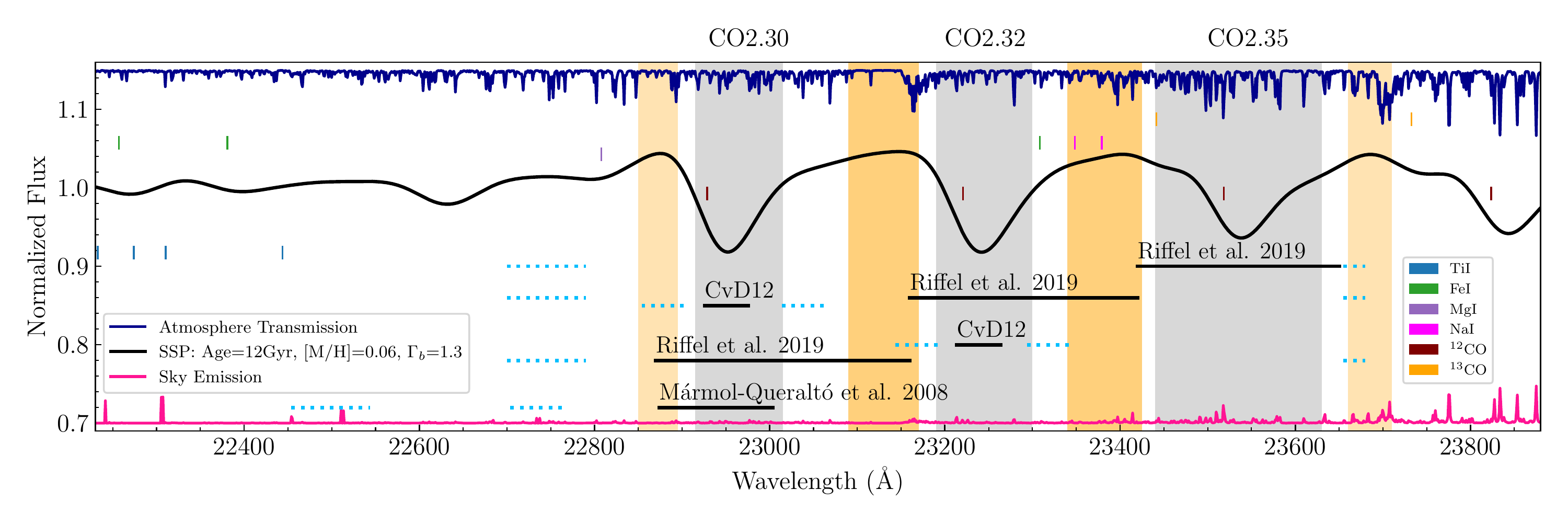}
    \caption{Same as Figure~(\ref{fig:figA1})}
    \label{fig:figA3}
\end{figure*}

\begin{landscape}
\begin{figure}
\centering
	\includegraphics[width=.99\linewidth]{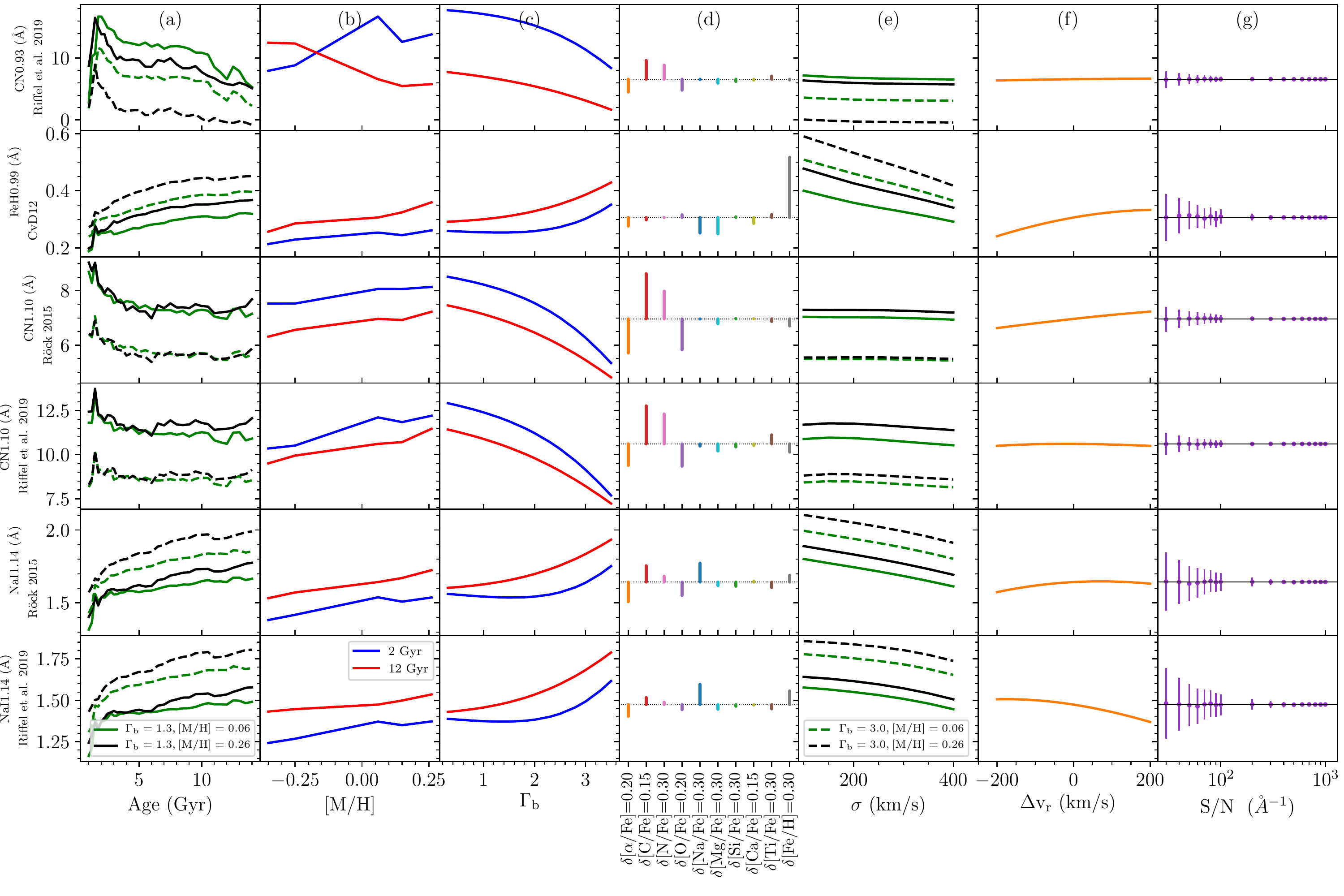}
    \caption{Model predictions of line-strength indices, defined in the literature, as a function of age (panels in Col. 'a'), metallicity  (Col. 'b'), IMF slope (Col. 'c'), elemental abundance ratios (Col. 'd'), velocity dispersion (Col. 'e'), radial velocity (Col. 'f') and SNR (Col. 'g'). All plots are based on E-MILES models with BaSTI evolutionary tracks, but those in Col. 'd', where we used CvD12 SSPs to estimate the effect of elemental abundance variations. (Panels 'a') Different colors correspond to different metallicity, i.e. [M/H] = +0.06 (green) and [M/H] = +0.26 (black). Models with a Kroupa-like IMF are shown as solid lines while models with a bottom-heavy IMF ($\Gamma_{b}=3.0$) are shown as dashed lines. (Panels 'b') The blue lines are model predictions for young populations with an age 2 Gyr while the red lines correspond to old (12 Gyr) populations. (Panels 'c') Blue and red lines are predictions for young (2 Gyr) and old (12 Gyr) populations, respectively. (Panels 'd') The vertical lines show the variation of a given line-strength for variations of different elemental abundances, [X/Fe]'s, shown with different colours. (Panels 'e') Different colors correspond to different metallicities, i.e. [M/H] = +0.06 (green) and [M/H] = +0.26 (black), respectively. Models with a Kroupa-like IMF are shown as solid lines while models with a bottom-heavy IMF ($\Gamma_{b}=3.0$) are shown as dashed lines. (Panels 'f') The orange line shows index measurements on a reference E-MILES SSP spectrum shifted to a given radial velocity, $\rm \Delta v_r$. (Panels 'g') The error bars show the average uncertainties on index values  as a function of S/N.
    }
    \label{fig:figA4}
\end{figure}
\end{landscape}

\begin{landscape}
\begin{figure}
\centering
	\includegraphics[width=\linewidth]{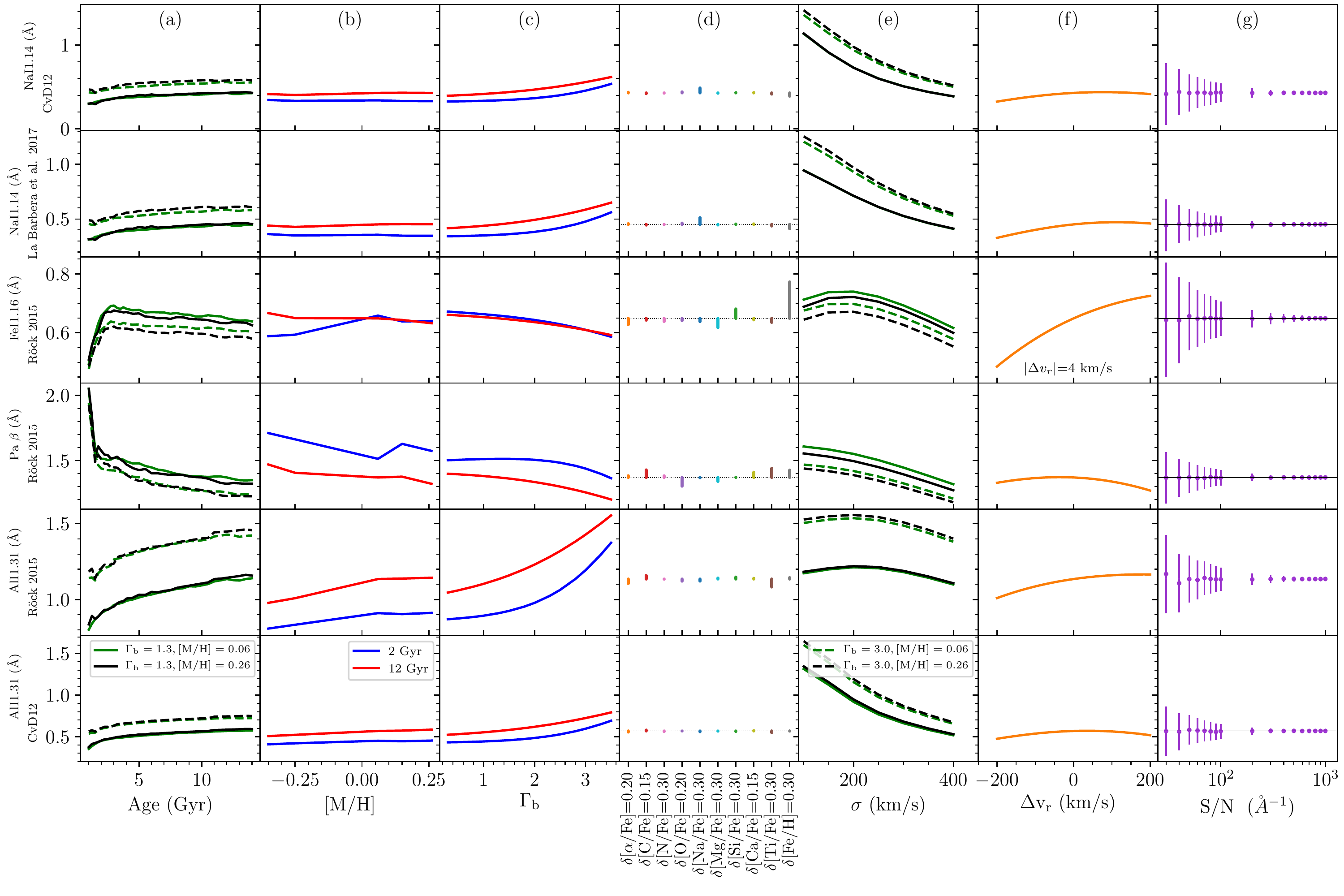}
    \caption{Same as Figure~(\ref{fig:figA4})}
    \label{fig:figA5}
\end{figure}
\end{landscape}

\begin{landscape}
\begin{figure}
\centering
	\includegraphics[width=\linewidth]{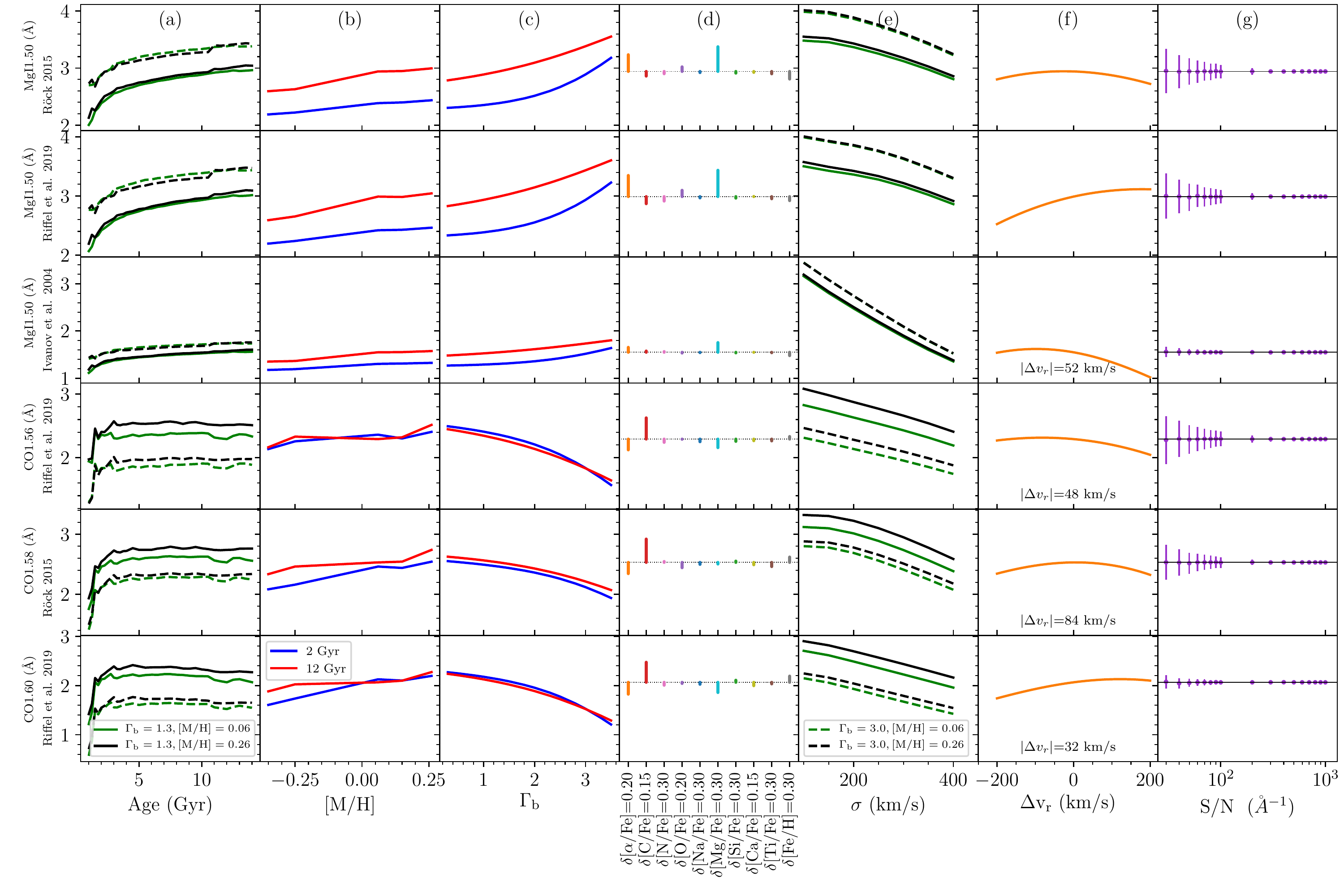}
    \caption{Same as Figure~(\ref{fig:figA4})}
    \label{fig:figA6}
\end{figure}
\end{landscape}

\begin{landscape}
\begin{figure}
\centering
	\includegraphics[width=\linewidth]{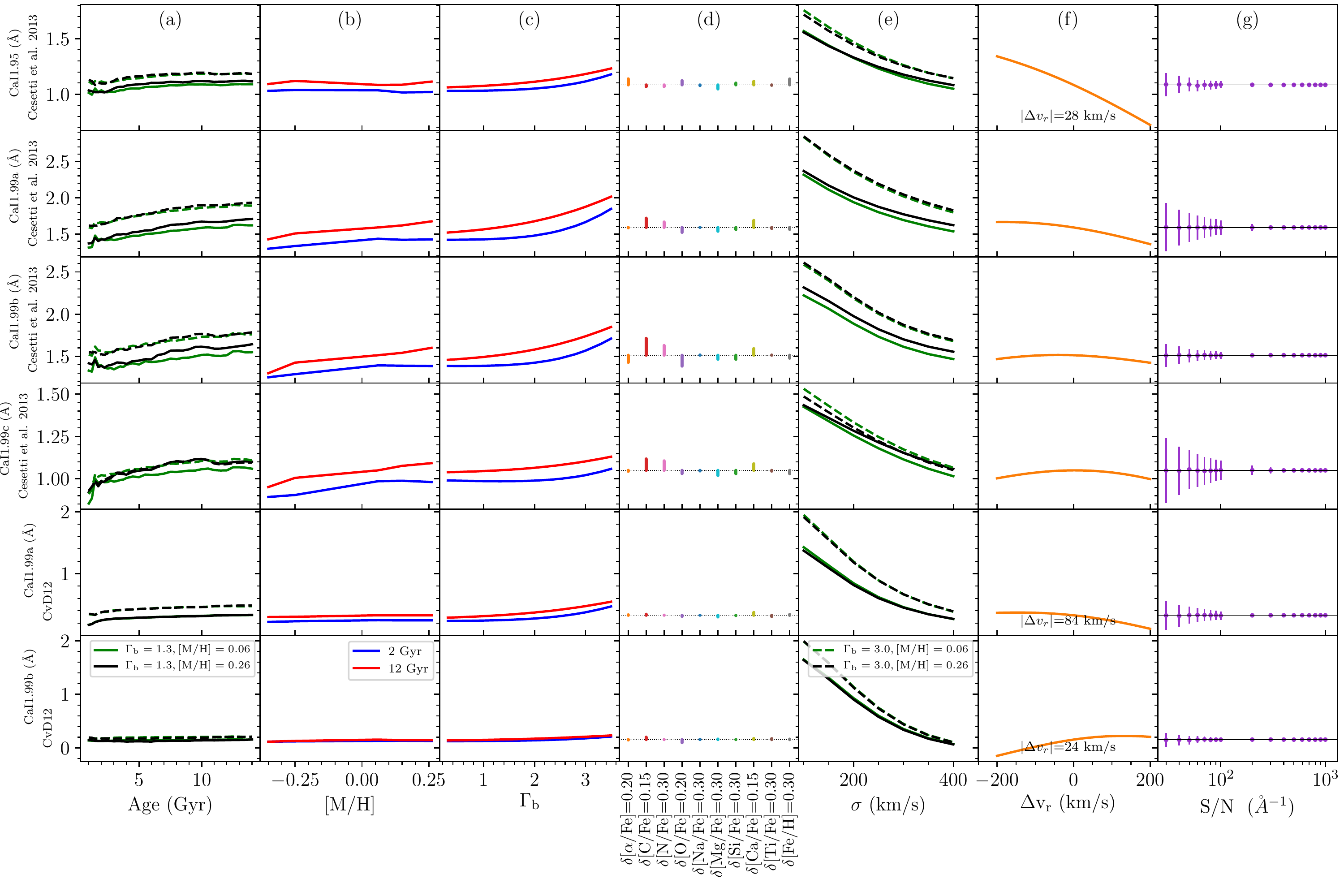}
    \caption{Same as Figure~(\ref{fig:figA4})}
    \label{fig:figA7}
\end{figure}
\end{landscape}

\begin{landscape}
\begin{figure}
\centering
	\includegraphics[width=\linewidth]{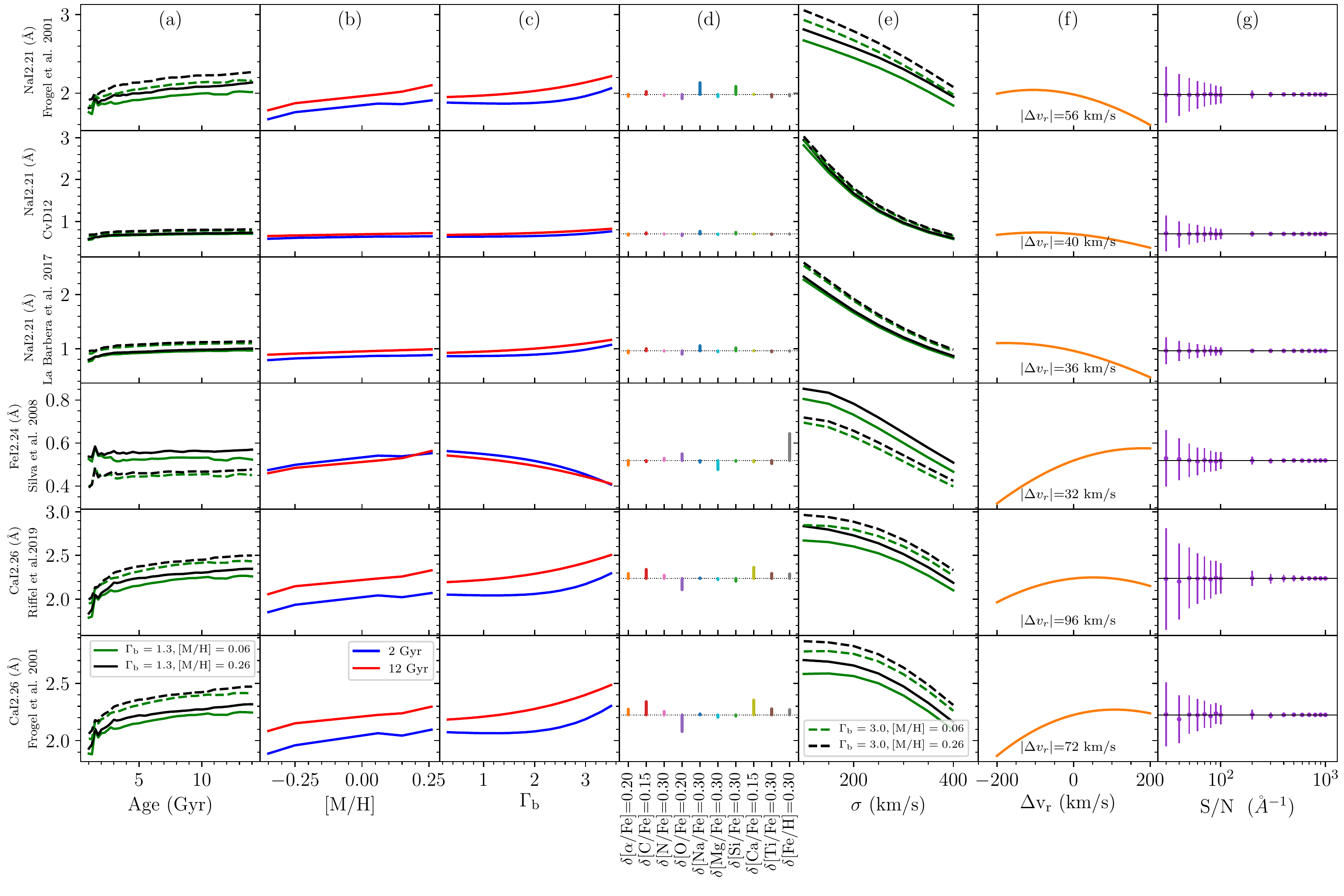}
    \caption{Same as Figure~(\ref{fig:figA4})}
    \label{fig:figA8}
\end{figure}
\end{landscape}

\begin{landscape}
\begin{figure}
\centering
	\includegraphics[width=\linewidth]{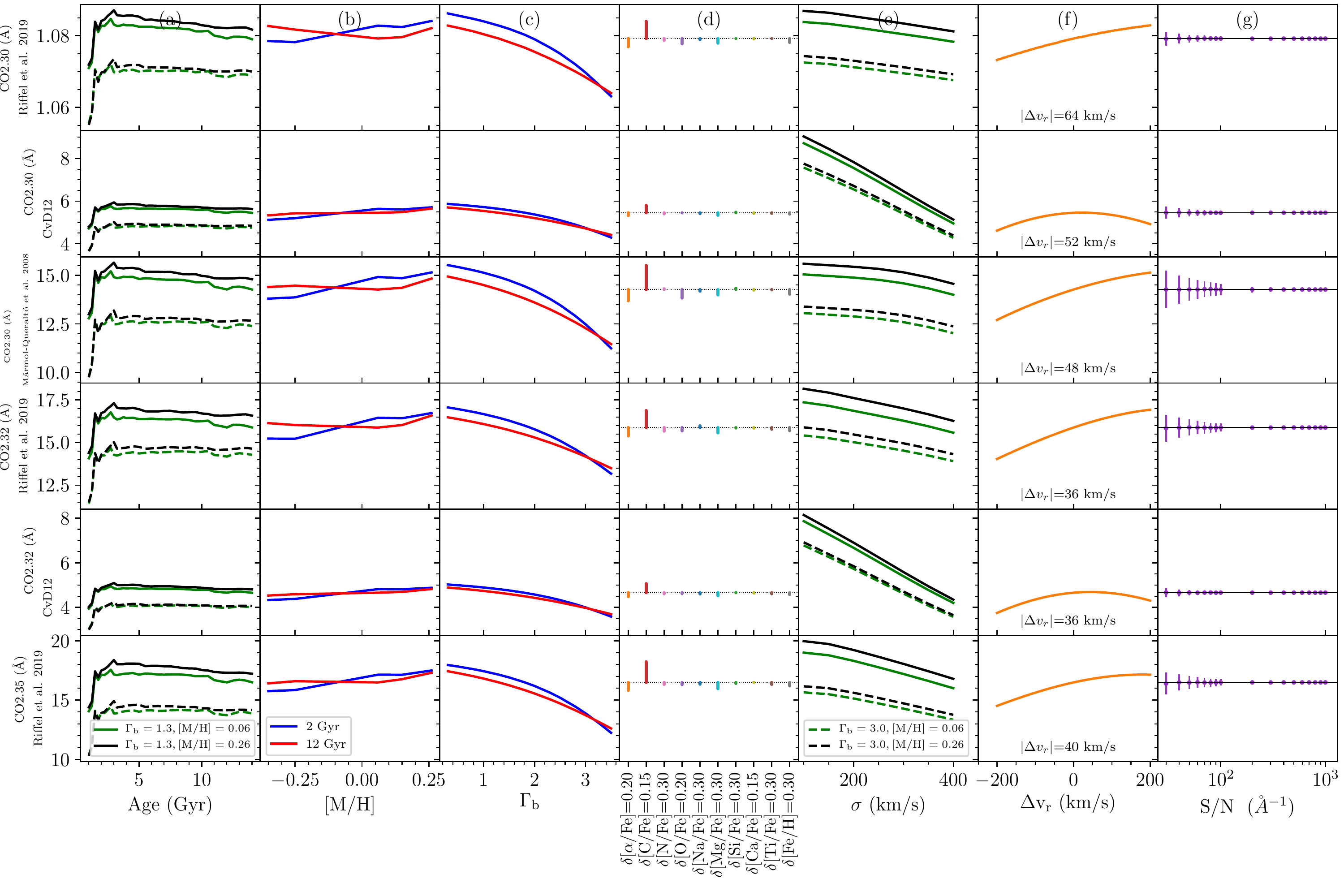}
    \caption{Same as Figure~(\ref{fig:figA4})}
    \label{fig:figA9}
\end{figure}
\end{landscape}

In Figs.~\ref{fig:figA1} to \ref{fig:figA3}, the spectrum of an E-MILES SSP, convolved to $\sigma=360$\kms, is shown in black. Sky emission lines and telluric absorption lines from Skycalc are shown in pink and dark blue, respectively. Our definition is indicated with a grey region for the central bandpass limits and in orange for pseudo-continua limits. We mark the definitions of the same feature in the literature by solid black lines for the central bandpasses and dotted blue lines as side bandpasses. The small vertical lines indicate the location of atomic and molecular absorptions identified in IRTF library, \citet{kleinmann1986} and \citet{lagattuta2017} (these are present in high-resolution spectra of cool stars). In Figs.~\ref{fig:figA4} to \ref{fig:figA9}, we compare the sensitivity of each index definition in the literature to stellar population parameters (age, metallicity and IMF slope), elemental abundance ratios, velocity dispersion, wavelength shifts and SNR. The spectral variations due to changes in individual elemental abundances are derived from CvD12 models.

Panel 'a' of Fig.~\ref{fig:figA1} shows two deep and broad features at the beginning of the \textit{J}-band ($\sim$9200 \AA \space and $\sim$9400 \AA). CN is the major source of continuum opacity in this wavelength range. This index is the most difficult one to define due to the presence of many atomic and molecular absorptions. Recently, \citet{riffel2019} provided a definition for this spectral region. Since the blue pseudo-continuum in their definition is located on top of a relatively strong absorption, we decided to re-define this index. The new index defined here place the blue-continuum in a region that is not blended by any atomic line. Also, we shifted the red-continuum to the blue to decrease the wavelength range covered by the index. We modified the central bandpass of the feature as well. 
In the same panel, we compare our definition for the FeH absorption to the one of CvD12. The small light blue vertical line shows the position of FeH line in the high-resolution spectrum of Arcturus. According, to this plot, the central bandpass limits in CvD12 definition, do not encompass this line. In our new definition we modified the feature bandpass to include it. Accordingly we changed the position of blue and red pseudo-continua to not overlap with the central bandpass limits. 

In panel 'b' of Fig.~\ref{fig:figA1}, the deep and broad absorption feature at $\sim$11000 \AA \space corresponds to the CN. Since continuum bands are far apart in \citet{rock2015phd} and \citet{riffel2019} definitions, we revised the definition of this index, putting the sidebands closer to the feature. It is noteworthy to mention that unlike \citet{riffel2019}'s definition, ours does not depend on the titanium abundance (comparing Col. 'd' of Fig.~\ref{fig:fig2} and Fig.~\ref{fig:figA4}). In the same panel, different definitions of the sodium absorption are shown. The sodium definition by \citet{labarbera2017} is local, and its central bandpass is almost identical to CvD12. However, both definitions are not optimized for velocity dispersion broadening. The blue continuum in \citet{riffel2019}'s definition is not far enough from the central bandpass and the one of \citet{rock2015phd} is placed on an absorption. In our new index, we made the central bandpass narrower, avoided sky emission at $\sim$11328 \AA \space and put the sidebands far enough from the feature. \ion{Fe}{i}1.16 is the next index, in this panel. It can be seen that our definition for iron absorption avoids the strong sky emission line around 11600 \AA. Index trend plots in the Fig.~\ref{fig:fig2} and Fig.~\ref{fig:figA5} show that our definition, compared to that of \citet{rock2015phd}, is more sensitive to IMF variations for both, young and old stellar populations. Moreover, the minimum SNR required for our definitions is much lower than that of \citet{rock2015phd}. \ion{Mg}{i}1.18 definition at panel 'b' of Fig.~\ref{fig:figA1} shows that two species contribute to the absorption feature: \ion{Mg}{i} and \ion{Ca}{ii}. While the red pseudo-continuum is free from absorption lines, the blue pseudo-continuum is contaminated by $\rm C_{2}$ line. According to Col. 'd' of Fig.~\ref{fig:fig3}, \ion{Ca}{ii} and $\rm C_{2}$ lines contamination do not affect the \ion{Mg}{i}1.18 index. In fact, enhancing the [Ca/Fe] and [C/Fe] does not change significantly the \ion{Mg}{i}1.18 line-strength.   

The wavelength limits of Pa $\beta$ index in panel 'c' of Fig.~\ref{fig:figA1} clearly show the advantage of our definition with respect to that of \citet{rock2015phd}; here the red pseudo-continuum is 150 \AA \space far from the feature. The relative more contribution of \ion{Ti}{i} lines, compared to \ion{Fe}{i} and \ion{Ca}{i} lines, in the absorption feature can explain the shallow dependence of this index on [Ti/Fe] variations (see Col. 'd' of Fig.~\ref{fig:fig3}). The red part of the same panel shows various definitions for the aluminium absorption feature. For the definition of \citet{rock2015phd}, the central bandpass is wider than CvD12 and its red sideband is much wider and extends further into the red. Its blue sideband also extends further into the blue compared to CvD12 and our definitions. Figure~\ref{fig:figA5} shows that the \ion{Al}{i}1.31 definition of CvD12 is strongly dependent on the velocity dispersion broadening as it does not leave any room for broadening. This figure also shows that the index of \citet{rock2015phd} requires a higher SNR than ours. Since our definition for aluminium does not include any \ion{Ti}{i} line, it is independent of the titanium abundance, unlike the index of \citet{rock2015phd}.

Panel 'd' of Fig.~\ref{fig:figA1} shows CN indices at the beginning of \textit{H}-band: one at $\sim$14200 \AA \space and the other one at $\sim$14600 \AA. They are located in a region with strong telluric absorptions and sky emissions. The central bandpass of CN1.46, includes contribution of VO molecule in addition to CN. Different definitions for magnesium absorption are shown in the same panel. The \ion{Mg}{i}1.50 feature has the central bandpass overlapping with  \citet{rock2015phd}, \citet{riffel2019} and \citet{ivanov2004} definitions, whereas the sidebandpasses are different. Our index avoids the spectral range covering \ion{Fe}{i} metallic lines. Moreover, we avoid strong emission lines in the pseudo-continuum ranges and the central bandpass. Comparing Fig.~\ref{fig:figA6} shows that our definition is more robust against $\rm \sigma$ compared to the \citet{ivanov2004} definition and requires significantly less SNR than \citet{rock2015phd} and \citet{riffel2019}'s ones. 

The CO1.56 index definition in panel 'a' of Fig.~\ref{fig:figA2} shows that the central bandpass and red pseudo-continuum of the CO index at $\sim$15600 \AA \space is contaminated by \ion{Fe}{i} lines. Moreover, \ion{Ti}{i}, \ion{Si}{i} and \ion{Ni}{i} are other contributors to this index. While definition of \citet{riffel2019} for CO includes \ion{Ti}{i} in red pseudo-continuum and \ion{Fe}{i} lines in blue-pseudo continuum, our definition is free of any lines in blue pseudo-continuum.  The CO1.58 index definition in the same panel, shows that \ion{Mg}{i} and \ion{Fe}{i} elements are other contributors of this index. As the red pseudo-continuum in definition of \citet{rock2015phd} is $\sim$300 \AA \space far away from the absorption, we redefined this index by placing the red pseudo-continuum close to the CO feature. Also note that the red pseudo-continuum in \citet{rock2015phd} definition includes \ion{Si}{i} and \ion{Ca}{i} lines, in addition to \ion{Fe}{i} lines, while in our definition, it is contaminated by \ion{Fe}{i} and \ion{Ti}{i} lines. None of these species causes a significant change on CO1.58 index as shown in Col. 'd' of Fig.~\ref{fig:fig4}. Definitions for CO index at $\sim$16000 \AA \space is shown in the same panel. Both definitions suffer from sky emission lines. However, the blue continuum in our index is in a region which is clean from telluric absorption lines. Moreover, since we put the blue continuum closer to the feature, the total spectral width of our CO1.60 index is smaller compared to the \citet{riffel2019}'s one.

Definitions of CO indices at $\sim$16400 \AA \space and $\sim$16600 \AA \space are shown in panel 'b' of Fig.~\ref{fig:figA2}. Both indices are contaminated by \ion{Fe}{i} lines but CO1.64 index is contaminated by \ion{Si}{i} line as well. They are located in a region that is not severely contaminated by telluric absorption lines. While for CO1.66 index we could avoid strong emission lines, the contamination of CO1.64 bandpasses was unavoidable. Definition for aluminium absorption at around 16700 \AA \space is shown in the same panel. While pseudo-continua are free from absorption lines, the central bandpass is contaminated by \ion{Fe}{i} lines. Note that this index is clean from telluric lines. There is only one weak telluric absorption in the feature bandpass and two sky emissions are present there as well. The bandpasses of CO index at around 16800 \AA \space is shown at the end of panel 'b' of Fig.~\ref{fig:figA2}. The index is located in region clean from telluric absorptions and our definitions avoids two strong sky emission lines at $\sim$16900 \AA \space and  $\sim$16950 \AA, respectively. The species that contaminate this index are \ion{Fe}{i}, \ion{Si}{i} and \ion{Ni}{i}. 

Panel 'c' of Fig.~\ref{fig:figA2} shows our definition for calcium absorption at around 19500 \AA \space and \citet{cesetti2013}'s definition. The central bandpass in the index of \citet{cesetti2013} is narrow, making it strongly dependant on the velocity dispersion broadening (see Col. 'e' of Fig.~\ref{fig:figA7}). Moreover, the pseudo-continua are far away from the feature. The same panel shows that our definition for \ion{Ca}{i}1.99a is different from that of \citet{conroy2012a} and \citet{cesetti2013}. The latter ones have wider central bandpass and wider red pseudo-continuum, which causes strong contamination from telluric absorption lines. However, our index is much less sensitive to $\rm \sigma$, specially in the lower mass regime according to Col. 'e' of Figs.~\ref{fig:fig5} and \ref{fig:figA7}. The comparison of our definition for \ion{Ca}{i}1.99b with those of CvD12 and  \citet{cesetti2013}, in the same panel, shows that our central bandpass  is  almost  identical to their definition although slightly wider, but our red sideband is much wider. According to Col. 'e' of Fig.~\ref{fig:figA7}, the CvD12 and \citet{cesetti2013} definitions suffer from a strong dependency on $\rm \sigma$. The comparison of our index and \citet{cesetti2013} definitions for \ion{Ca}{i}1.99c (the same panel) shows that our central bandpass is wider and the blue continuum is closer to the feature.

Panel 'd' of Fig.~\ref{fig:figA2} shows different definitions of the same feature, \ion{Na}{i}2.21. Unlike the indices of CvD12 and \citet{labarbera2017}, we left free some space between pseudo-continua and the feature bandpasses to minimise the effects of velocity dispersion broadening as shown by their dependence against $\rm \sigma$ (Col. 'e' of Figs.~\ref{fig:fig5} and \ref{fig:figA8}). Our definition for iron absorption at $\sim$22400 \AA \space is compared with \citet{silva2008} in the same panel. The blue pseudo-continuum in our definition is not blended with strong absorption line. Moreover, in our index the pseudo-continua are significantly closer to the feature bandpass, while the latter is wider. In the red part of panel 'd' of Fig.~\ref{fig:figA2}, we defined a modified index around the \ion{Ca}{i}. This index is slightly different from those used by \citet{frogel2001} and \citet{riffel2019}. Our red continuum is far enough from the central bandpass, making the index line-strength rather insensitive to velocity dispersion broadening. The new \ion{Ca}{i}2.26 definition is promising as an IMF-sensitive feature as, unlike the other definitions in the literature is less sensitive to the titanium, carbon and oxygen abundance (see Col. 'd' of Fig.~\ref{fig:figA8}). This shows that our index is less contaminated from neighbouring strong CO absorptions. 

Figure~\ref{fig:figA3} shows different definitions for the first bandhead of CO in the \textit{K}-band. The definition of \citet{marmol2008} has two blue bandpasses. The red bandpass of \citet{riffel2019} is far away from the absorption and is very narrow while CvD12 spectral index is local and its central bandpass is almost identical to our definition, although narrower. Finally, compared to the CvD12, our index is more robust against $\rm \sigma$ (see Col. 'e' of Figs.~\ref{fig:fig6} and \ref{fig:figA9}). In the same figure, we compared different definitions of CO2.32 in the literature as well. Unlike the index of \citet{riffel2019}, our definition is local, and only \ion{Na}{i} is within the red continuum. We carefully placed the bandpasses to minimise dependence on velocity dispersion. Hence, compared to the CvD12, our definition is more robust against $\rm \sigma$, specially for the low and intermediate-mass galaxy regime (see Col. 'e' of Figs.~\ref{fig:fig6} and \ref{fig:figA9}). Finally, this figure compares our definition for CO at $\sim$23540 \AA \space with \citet{riffel2019}'s definition. Our refined CO2.35 index has the central bandpass overlapping with their definition, whereas the continuum bandpasses are different. The blue bandpass of \citet{riffel2019} is about 400 \AA \space far from the absorption, and its red bandpass is narrow and touches the central bandpass, leaving no space for velocity dispersion broadening. Moreover, the required SNR for our definition is much lower than for their definition (see Col. 'g' of  Figs.~\ref{fig:fig6} and \ref{fig:figA9}).

\section{Uncertainties Induced by Removing Bad Pixels within the Index Definition } \label{sec:appendixB}

In the following, we show how index line-strengths are affected by the masking of a given fraction of pixels within each bandpass, translating these variations into changes of the main stellar population parameters involved in a stellar population analysis, such as age, metallicity, and IMF. In Table~\ref{tab:tabB1}, we also provide, as a quick reference for the user, a more quantitative estimate of the maximum fraction of pixels with substantial systematic uncertainties that is allowed to perform a reliable stellar population analysis. We also provide below, a practical example showing how a user may benefit from the numbers provided, for each index, in Table~\ref{tab:tabB1}.

\begin{figure*}
	\includegraphics[width=.9\linewidth]{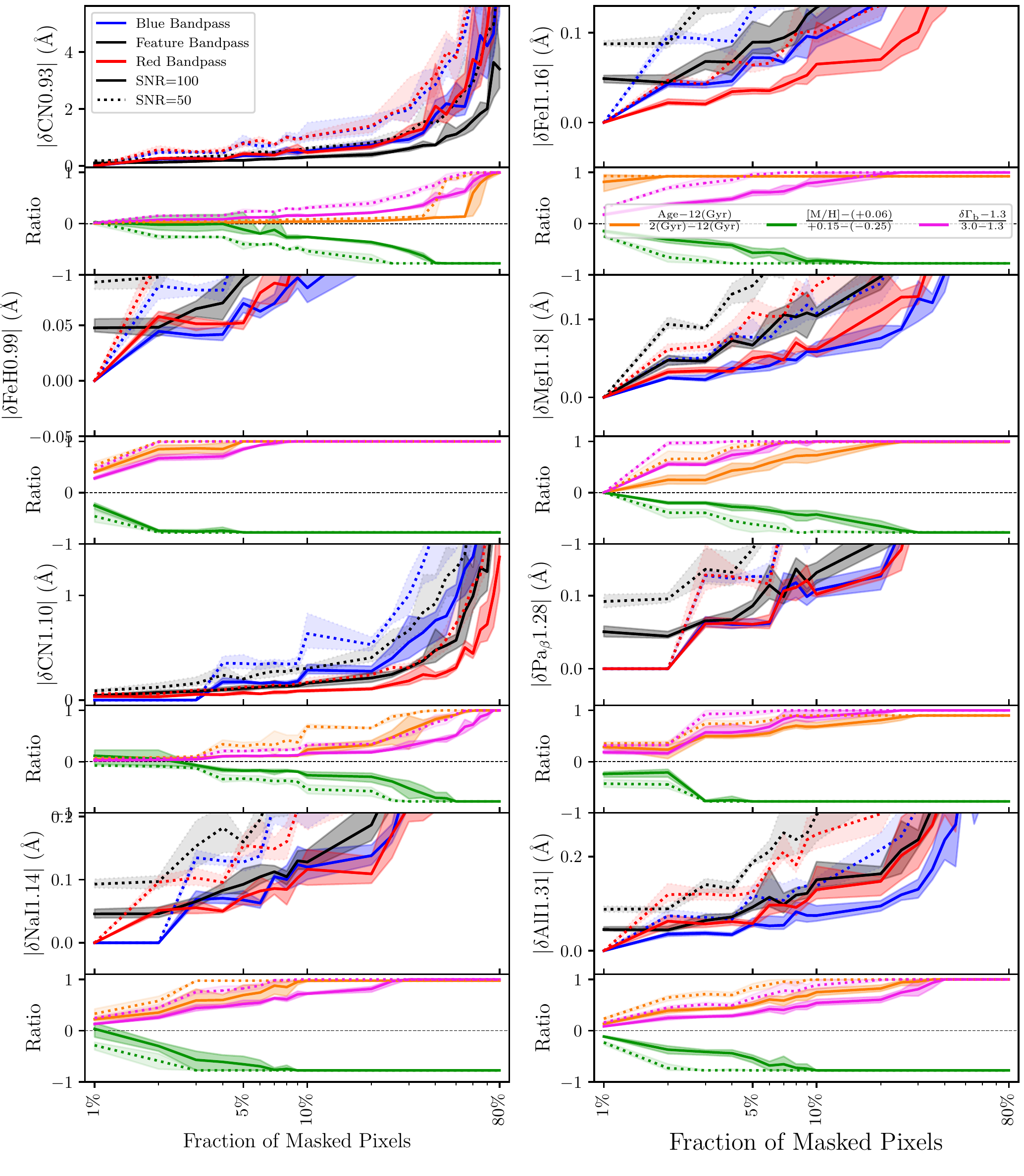}
    \caption{Effect of removing a given fraction of pixels within blue (blue lines), central (black lines) and red (red lines) bandpasses on the line-strength of NIR indices (upper panels). Solid and dotted lines correspond to  spectra with SNR=100 and 50 \AA$^{-1}$, respectively. 
    In the lower panels, the changes of index values plotted in the upper panels have been transformed into relative variations of stellar population parameters, i.e. age, metallicity, and IMF slope.
    The changes in stellar population parameters are computed with respect to a reference spectrum with age 12 Gyr, [M/H]=0.06, and $\Gamma_{b}=1.3$, and are normalized to the range of stellar populations parameters of intermediate-mass and massive ETGs. Shaded areas, in both the upper and lower panels, show the effect of varying velocity dispersion from 60 to 360 \kms.}
    
    \label{fig:figB1}
\end{figure*}

\begin{figure*}
	\includegraphics[width=.95\linewidth]{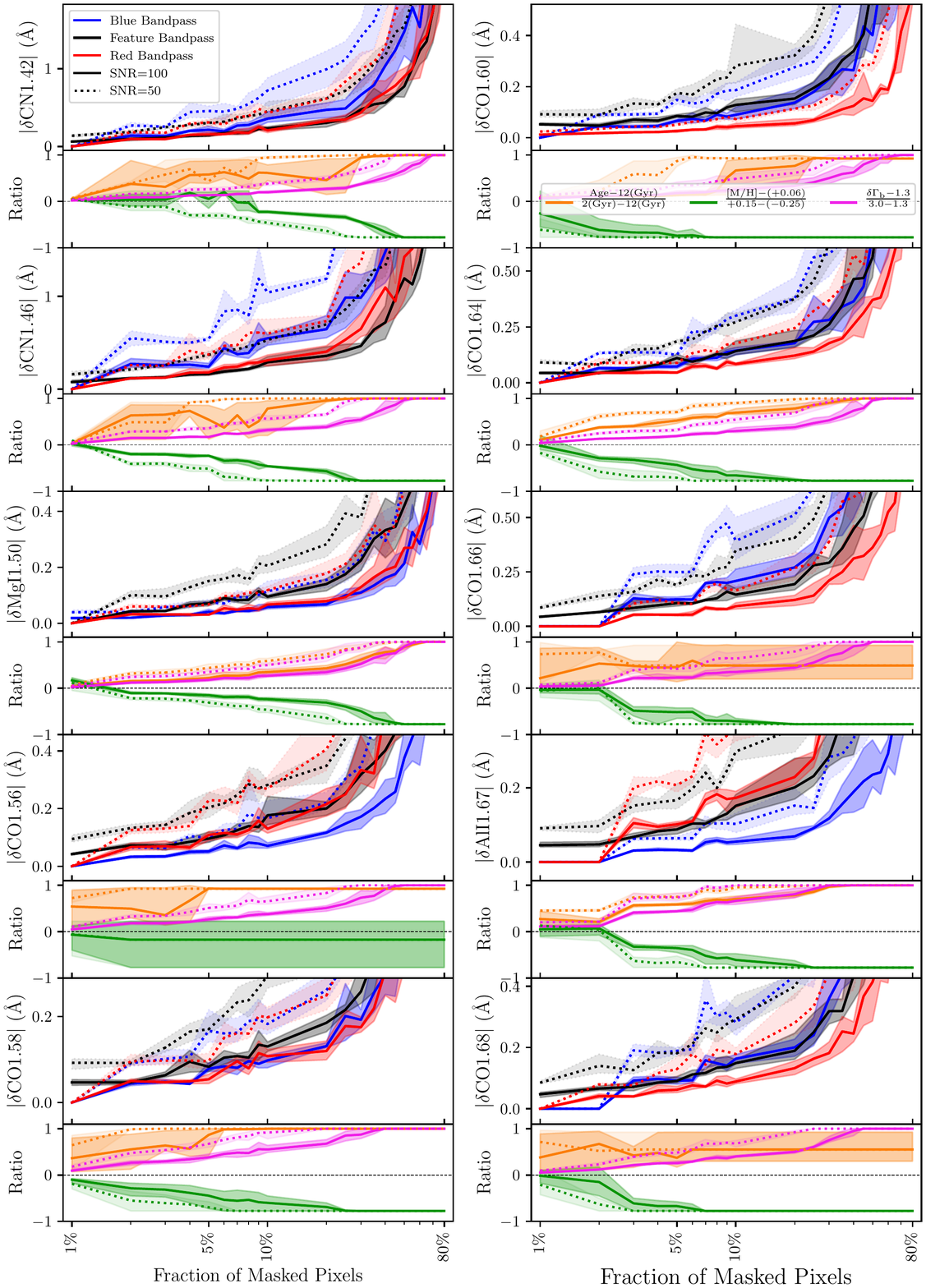}
    \caption{Same as Figure~(\ref{fig:figB1})}
    \label{fig:figB2}
\end{figure*}

\begin{figure*}
	\includegraphics[width=.95\linewidth]{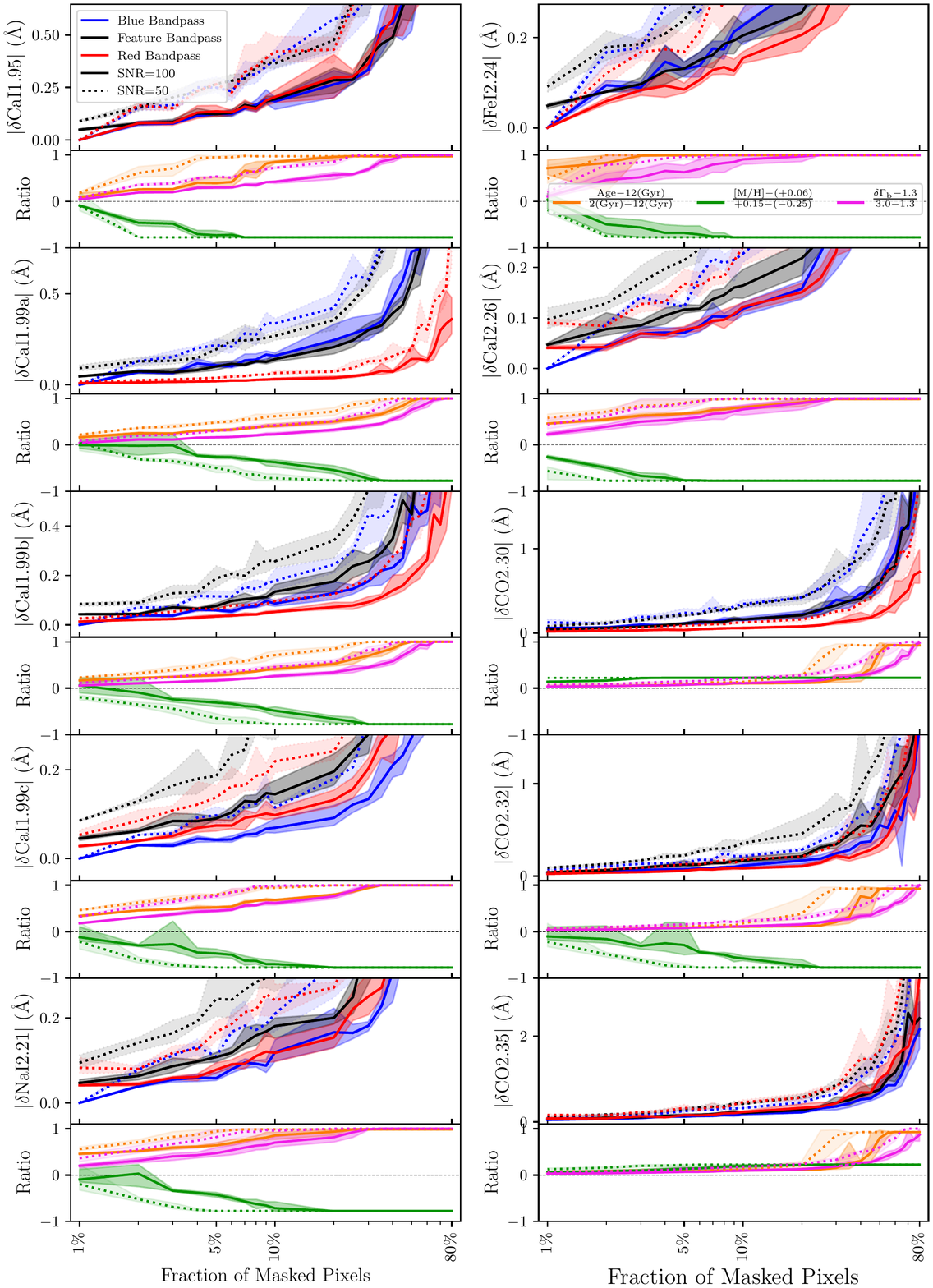}
    \caption{Same as Figure~(\ref{fig:figB1})}
    \label{fig:figB3}
\end{figure*}

Figs.~\ref{fig:figB1} to \ref{fig:figB3} show, for all the new indices in the present work, the variation of index line-strengths when removing bad pixels within an index definition   (see upper panels). In the same Figures, we also translate the above variation into relative changes of the relevant stellar population properties, plottting these relative changes as a function of the fraction of masked pixels (see lower panels). To this effect, as already detailed in Sect.~\ref{sec:reliability}, we used an E-MILES SSP with age 12 Gyr, [M/H] = +0.06 dex, and $\Gamma_{b}=1.3$ as our reference model and randomly removed a percentage of the total pixels within each bandpass defining a given index, before measuring its strength. We repeated this procedure 1000 times and saved the maximum change in the index strength. For each index, in corresponding upper panel, the changes in the index value as a function of the fraction of removed pixels (x-axis) is shown, with solid lines for spectra having SNR=100 \AA$^{-1}$ and dotted lines for spectra with SNR=50 \AA$^{-1}$. Blue, black and red colours correspond to removing a given percentage of pixels within the blue, feature, and red bandpasses of the index, respectively. To assess the impact of spectral resolution on the minimum fraction of pixels required to measure an index, we show the results obtained with different resolutions, from 60 to 360 \kms, as shaded regions. In the lower panels of the Figures, for each index, we translated the average change of index value due to the masking of a given fraction of pixels within the index bandpasses to relative changes in the main stellar population parameters, i.e. age, metallicity, and IMF. The orange, green, and pink lines show relative changes in age, metallicity, and IMF slope (with respect to our reference model), respectively.  For each stellar population parameter, we normalize variations by a range of value for that parameter typical for intermediate- and high-mass ETGs (i.e. age range from 2 to 12 Gyr, total metallicity from +0.15 to -0.25~dex, and IMF variation from Kroupa-like ($\Gamma_{b}=1.3$) to bottom-heavy ($\Gamma_{b}=3.0$)) (see, e.g., \citealt{labarbera2013}). As for the upper panels, the solid and dotted lines show the results for spectra with SNR=100 \AA$^{-1}$ and SNR=50 \AA$^{-1}$, respectively, while
the impact of spectral resolution, from 60 to 360 \kms, is shown with shaded area. Note that, in these plots, the upper limits are imposed by the maximum variation on the index value due to the typical stellar population variation of intermediate-mass and massive ETGs (i.e. we do not consider relative variations larger than one).

In principle, the plots in Figs.~\ref{fig:figB1} to \ref{fig:figB3} should be used to assess the robustness of spectral indices against systematic uncertainties for a given galaxy sample/spectroscopic dataset. In order to facilitate the use of these plots, we have also defined a maximum fraction of masked pixels that can be allowed to use each index. We defined this fraction as the one for which the ratios of stellar population parameters changes do not exceed $\pm$0.5. Results are reported in Table~\ref{tab:tabB1}. Note that as mentioned in Sect.~\ref{sec:reliability} this table should be only intended as a quick reference for the user, while different criteria can be defined based on a specific application. Column 1 in Table~\ref{tab:tabB1} indicates the name of indices, with SNR being given in Col. 2. Columns 3, 4, and 5 provide the maximum fraction of pixels which can be removed within the index bandpasses and do not impact our ability to derive the age, metallicty, and IMF. 

\strutlongstacks{T}
\begin{table}
	\centering
	\caption{Maximum fraction of bad pixels within bandpass limits of spectral indices, for which it can be derived meaningful stellar population parameters. Column 1 gives the name of the indices. Column 2 indicates the SNR of the reference spectrum used to derive the fractions. Columns 3, 4, and 5 correspond to the maximum fractions of bad pixels that can be masked out within the index bandpasses
	to derive stellar population parameters.}
	\label{tab:tabB1}
	\begin{tabular}{lcccr } 
		\hline\hline
		Index & SNR	& Age	& Metallicity	& IMF 	\\
		  &  (\AA$^{-1})$	& ($\%$)	& ($\%$)	& ($\%$) 	\\
		(1) & (2)	& (3)	& (4)	& (5)  \\
		\hline
CN0.93 & 100 & 55 & 25 & 40 \\
CN0.93 & 50 & 40 & 7 & 20 \\
FeH0.99 & 100 & 2 & 2 & 2 \\
FeH0.99 & 50 & - & 2 & 2 \\
CN1.10 & 100 & 25 & 25 & 45 \\
CN1.10 & 50 & 9 & 9 & 25 \\
NaI1.14 & 100 & 2 & 2 & 4 \\
NaI1.14 & 50 & 2 & 2 & 2 \\
FeI1.16 & 100 & - & 4 & 4 \\
FeI1.16 & 50 & - & 2 & 2 \\
MgI1.18 & 100 & 5 & 10 & 2 \\
MgI1.18 & 50 & 2 & 3 & 2 \\
Pab1.28 & 100 & 4 & 2 & 2 \\
Pab1.28 & 50 & 2 & 2 & 2 \\
AlI1.31 & 100 & 4 & 4 & 9 \\
AlI1.31 & 50 & 2 & 2 & 2 \\
CN1.42 & 100 & 3 & 30 & 35 \\
CN1.42 & 50 & 2 & 10 & 10 \\
CN1.46 & 100 & 2 & 10 & 20 \\
CN1.46 & 50 & 3 & 4 & 8 \\
MgI1.50 & 100 & 20 & 30 & 25 \\
MgI1.50 & 50 & 6 & 10 & 10 \\
CO1.56 & 100 & 3 & - & 10 \\
CO1.56 & 50 & - & - & 4 \\
CO1.58 & 100 & 2 & 5 & 8 \\
CO1.58 & 50 & - & 2 & 2 \\
CO1.60 & 100 & 8 & 2 & 25 \\
CO1.60 & 50 & 2 & - & 10 \\
CO1.64 & 100 & 4 & 5 & 25 \\
CO1.64 & 50 & 2 & 2 & 8 \\
CO1.66 & 100 & 5 & 3 & 20 \\
CO1.66 & 50 & - & 2 & 6 \\
AlI1.67 & 100 & 2 & 6 & 6 \\
AlI1.67 & 50 & 2 & 2 & 2 \\
CO1.68 & 100 & 5 & 2 & 10 \\
CO1.68 & 50 & - & 2 & 5 \\
CaI1.95 & 100 & 6 & 3 & 10 \\
CaI1.95 & 50 & 2 & 2 & 4 \\
CaI1.99a & 100 & 10 & 10 & 35 \\
CaI1.99a & 50 & 5 & 5 & 10 \\
CaI1.99b & 100 & 10 & 10 & 30 \\
CaI1.99b & 50 & 4 & 3 & 10 \\
CaI1.99c & 100 & 3 & 6 & 6 \\
CaI1.99c & 50 & 2 & 2 & 2 \\
NaI2.21 & 100 & 2 & 6 & 5 \\
NaI2.21 & 50 & - & 2 & 2 \\
FeI2.24 & 100 & - & 2 & 2 \\
FeI2.24 & 50 & 2 & 2 & 2 \\
CaI2.26 & 100 & 2 & 2 & 3 \\
CaI2.26 & 50 & - & - & 2 \\
CO2.30 & 100 & 45 & - & 55 \\
CO2.30 & 50 & 20 & - & 35 \\
CO2.32 & 100 & 35 & 8 & 50 \\
CO2.32 & 50 & 20 & 2 & 35 \\
CO2.35 & 100 & 45 & - & 60 \\
CO2.35 & 50 & 20 & - & 35 \\
		\hline
	\end{tabular}
\end{table}

As an example, we discuss the reliability of our index measurements for the sample of \cite{labarbera2019} (see Sect.~\ref{sec:example_applications}). For each galaxy, we identified bad pixels within each bandpass of the NIR indices. The fractions are reported in Table~\ref{tab:tabB2}. Since we have analyzed how \ion{Mg}{i}1.18 and \ion{Mg}{i}1.50 can help constraining the stellar IMF (see Fig.~\ref{fig:fig9}),
we compare the estimated fraction of affected pixels with those provided in Column 5 of  Table~\ref{tab:tabB1}. 
The SNR of the spectra is high ($>100$ \AA$^{-1}$); therefore, we consider the percentages for SNR = 100 \AA$^{-1}$. In case of \ion{Mg}{i}1.18, for a spectrum with SNR=100 \AA$^{-1}$, the maximum fraction of affected pixels to constrain the IMF is (according to our criterium) 2$\%$, within the bandpasses. According to Table~\ref{tab:tabB2}, for XSG2, XSG6, XSG7, and XSG10, the fraction of affected pixels for \ion{Mg}{i}1.18 does not exceed the maximum fraction in Table~\ref{tab:tabB1}. This shows that, to perform an IMF analysis, \ion{Mg}{i}1.18 index measurements are reliable for these galaxies. On the other hand, this is not the case for XSG1, XSG8, and XSG9, and therefore they were not included in the upper panels of Fig.~\ref{fig:fig9}. \ion{Ca}{i}1.95 index is reliable for analysing age and IMF slope for all galaxies in the sample but for XSG6, while it cannot be used to constrain metallicity for XSG6, XSG7, XSG9, and XSG10. \ion{Ca}{i}1.99a, \ion{Ca}{i}1.99b, and \ion{Ca}{i}1.99c indices fall within a clean region and none of the pixels are affected by telluric absorption or sky emission. However, for XSG8, we find that 11$\%$ of pixels in the blue bandpasses of these indices are contaminated, and thus one cannot rely on these indices to constrain metallicity of XSG8. In the case of \ion{Ca}{i}2.26, only for XSG6 one cannot constrain stellar population parameters, as the fraction of affected pixels within the red bandpass (5$\%$) is higher than the maximum fractions provided in Table~\ref{tab:tabB1} (2$\%$ for age and metallicity and 3$\%$ for IMF).

\strutlongstacks{T}
\begin{table*}
	\centering
	\caption{Fraction of pixels with substantial systematic uncertainties within the bandpasses of targeted indices in the spectra of \citet{labarbera2019} sample. Column 1 gives the name of indices. Column 2 indicates the bandpass of indices. The identified fraction of affected pixels is reported in Cols. 3 to 9 for each galaxy in the sample.}
	\label{tab:tabB2}
	\begin{tabular}{lcccccccr} 
		\hline\hline
		
		Index & Bandpass & XSG1 & XSG2	& XSG6	& XSG7 & XSG8 & XSG9 & XSG10 \\
		& & ($\%$)	& ($\%$)	& ($\%$) &	($\%$)	& ($\%$) & 	($\%$)	& ($\%$) \\
		(1) & (2)	& (3)	& (4)	& (5) &	(6)	& (7)	& (8)	& (9)  \\
	    \hline
\ion{Mg}{i}1.18 & blue & 3 & 0 & 0 & 2 & 3 & 2 & 0 \\
& feature & 0 & 0 & 0 & 0 & 0 & 0 & 0 \\
& red & 8 & 2 & 0 & 2 & 0 & 4 & 0 \\
       
\ion{Mg}{i}1.50 & blue & 0 & 0 & 1 & 0 & 0 & 0 & 7 \\
& feature & 0 & 0 & 0 & 16 & 10 & 19 & 0 \\
& red & 0 & 0 & 0 & 0 & 0 & 0 & 25 \\
        
\ion{Ca}{i}1.95 & blue & 0 & 0 & 5 & 0 & 0 & 2 & 5 \\
& feature & 2 & 0 & 5 & 5 & 2 & 5 & 5 \\
& red & 0 & 0 & 13 & 0 & 3 & 3 & 0 \\

\ion{Ca}{i}1.99a & blue & 0 & 0 & 0 & 0 & 11 & 0 & 0 \\
& feature & 0 & 0 & 0 & 0 & 0 & 0 & 0 \\
& red & 0 & 0 & 0 & 0 & 0 & 0 & 0 \\

\ion{Ca}{i}1.99b & blue & 0 & 0 & 0 & 0 & 11 & 0 & 0 \\
& feature & 0 & 0 & 0 & 0 & 0 & 0 & 0 \\
& red & 0 & 0 & 0 & 0 & 0 & 0 & 0 \\

\ion{Ca}{i}1.99c & blue & 0 & 0 & 0 & 0 & 11 & 0 & 0 \\
& feature & 0 & 0 & 0 & 0 & 0 & 0 & 0 \\
& red & 0 & 0 & 0 & 0 & 0 & 0 & 0 \\

\ion{Ca}{i}2.26 & blue & 0 & 0 & 0 & 0 & 0 & 0 & 0 \\
& feature & 0 & 0 & 1 & 1 & 0 & 0 & 0 \\
& red & 0 & 0 & 5 & 0 & 0 & 0 & 0 \\
		\hline
	\end{tabular}
\end{table*}


\bsp	
\label{lastpage}
\end{document}